\newcommand{\codename}{\textbf{ACID}}
\documentclass[fleqn,usenatbib]{mnras}

\usepackage{newtxtext,newtxmath}
\usepackage{float}
\usepackage{tabularx}
\usepackage[T1]{fontenc}
\usepackage{float}

\DeclareRobustCommand{\VAN}[3]{#2}
\let\VANthebibliography\thebibliography
\def\thebibliography{\DeclareRobustCommand{\VAN}[3]{##3}\VANthebibliography}

\usepackage{graphicx}	
\usepackage{amsmath}	

\usepackage{xcolor}
\usepackage{hyperref}
\usepackage{gensymb}
\newcommand{\orc}{\includegraphics[height=\fontcharht\font`A]{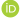}}
\newcommand{\orcid}[1]{\href{https://orcid.org/#1}{\orc}}

\title[A.C.I.D]{A.C.I.D - An Improved LSD Technique for Accurate Line Profile Retrieval.}

\author[L. Dolan et al.]{
L. S. Dolan,\orcid{0000-0003-4260-7225}\thanks{E-mail: ldolan05@qub.ac.uk}$^{1}$
E. J. W. de Mooij,$^{1}$\orcid{0000-0001-6391-9266}
C. A. Watson$^{1}$
and D. G. Jackson\orcid{0000-0003-0343-7905}$^{1}$\\
$^{1}${Astrophysics Research Centre, School of Mathematics \& Physics, Queen’s University Belfast, University Road, Belfast BT7 1NN, UK}
}

\date{Accepted XXX. Received YYY; in original form ZZZ}

\pubyear{2023}

\begin{document}
\label{firstpage}
\pagerange{\pageref{firstpage}--\pageref{lastpage}}
\maketitle

\begin{abstract}
Stellar activity and planetary effects induce radial velocity (RV) offsets and cause temporal distortions in the shape of the stellar line profile. Hence, accurately probing the stellar line profile offers a wealth of information on both the star itself and any orbiting planets. Typically, Cross-Correlation Functions (CCFs) are used as a proxy for the stellar line profile. The shape of CCFs, however, can be distorted by line blending and aliasing limiting the stellar and planetary physics that can be probed from them. Least-squares deconvolution (LSD) offers an alternative that directly fits the mean line profile of the spectrum to produce a high-precision profile. In this paper, we introduce our novel method $\codename$ (\textbf{A}ccurate \textbf{C}ontinuum f\textbf{I}tting and \textbf{D}econvolution) that builds on LSD techniques by simultaneously fitting the spectral continuum and line profile as well as performing LSD in effective optical depth. Tests on model data revealed $\codename$ can accurately identify and correct the spectral continuum to retrieve an injected line profile. $\codename$ was also applied to archival HARPS data obtained during the transit of HD189733b. The application of the Reloaded Rossiter-McLaughlin technique to both $\codename$ profiles and HARPS CCFs shows $\codename$ residual profiles improved the out-of-line RMS by over $5\%$ compared to CCFs.
Furthermore, $\codename$ profiles are shown to exhibit a Voigt profile shape that better describes the expected profile shape of the stellar line profile. This improved representation shows that $\codename$ better preserves the stellar and planetary physics encoded in the stellar line profile shape for slow rotating stars.
\end{abstract}




\section{Introduction}

\indent Stellar line profiles offer a wealth of planetary and stellar information that can be obtained from spectroscopic data. For example, stellar radial velocities (RVs), measured from velocity shifts of stellar line profiles, are used in the detection of exoplanets and, when combined with photometric observations, to extract bulk planetary properties \citep[e.g.][]{2023MNRAS.522..750D, 2023MNRAS.518.4845J}.
\newline \indent RV measurements can also be used to detect the anomalous radial velocity shifts that occur when a planet transits in front of a rotating star, known as the classical Rossiter-McLaughlin (RM) effect. This can probe the orbital alignment of an exoplanet system \citep{1924ApJ....60...15R, 1924ApJ....60...22M} and allows us to differentiate between planetary migration models as more `violent' migration mechanisms such as the Kozai-Lidov mechanism \citep{1962AJ.....67..591K, 1962P&SS....9..719L}, third body interactions \citep{2007ApJ...669.1298F, 2003ApJ...585.1024G} and planet-planet scattering are expected to produce systems with misaligned or inclined orbits. The classical RM effect has been routinely used for a range of different systems \citep[e.g.][]{2014A&A...564L..13E, 2022A&A...668A.172G} and hence improved our understanding of planetary formation and evolution. Precise RV measurements induced by planetary effects such as classical RM are generally limited to very stable instruments such as the High Accuracy Radial-velocity Planet Searcher (HARPS). Line-profile tomography measures the temporal change in the shape of line profile induced by the RM effect \citep{2010MNRAS.403..151C}. As a planet transits in front of a star the planet casts a `Doppler shadow', manifesting as an identifiable bump in the line profile. Modelling this bump throughout a transit allows the RM effect to be measured regardless of the stability of the instrument. In the case of stable instruments, this technique can better constrain planetary and stellar parameters as both the RV shift and distortion in the shape of the line profile can be measured.
\newline \indent As well as obtaining planetary parameters the stellar line profile can also contribute to our understanding of stellar activity. Cool starspots distort the line profile as the spot passes across the visible stellar surface. In this case, the cooler and therefore dimmer spot on the stellar surface manifests as a bump in the line profile, similar to that which occurs during a planet's transit \citep{1983PASP...95..565V}. Modelling this distortion can provide information on spot coverage on the star as well as the brightness and temperature of the spot \citep{1997astro.ph..4191V}. Measuring the temporal evolution of the bump provides an estimate for the stellar rotational velocity and even the rate of differential rotation in the case of multiple spots that stretch across different stellar latitudes \citep{1992A&A...259..183S, 2003A&A...410..685O}. Other manifestations of stellar magnetic activity, such as bright plage regions, also impact the line profiles and stellar activity can affect the width and asymmetry of different lines in different ways \citep{2017MNRAS.468L..16T}. These activity-induced line profile shape variations currently pose a major challenge to the detection and characterisation of exoplanets, especially as the field progresses to detect Earth-analog planets. Indeed, incorrect treatment of stellar activity has led to false planetary detections and so it is vital to better understand stellar activity's role in the distortion of the line profile (and to be able to account for it) in the context of the discovery and characterisation of exoplanets \citep{2014ExA....38..249R,2022AJ....163..215S}.
\newline \indent Many of these astrophysical studies mentioned rely on detecting sub-m/s radial velocity variations and equivalently small changes to the stellar line profile shape. Hence, obtaining accurate and precise measurements of stellar line profiles and associated information is vital in correctly detecting and interpreting these subtle effects. To achieve this, often the information from many thousand line profiles must be combined to boost the effective signal-to-noise (S/N). The quality and accuracy of these methods will be highly dependent on the technique used. For example, Cross-Correlation Functions (CCFs) are commonly used as proxy stellar line profiles and to extract stellar line profile information. A CCF is generally constructed by cross-correlating a stellar spectrum with a model spectrum and thus measures the similarity between them. Alternatively, as with HARPS, the CCFs are constructed by cross-correlating with a line mask, outlining the wavelength regions containing the expected stellar lines \citep{exoplanet_handbook_2018}. The final profile is equivalent to the sum of all the lines, taken directly from the spectrum and identified by the mask. In this case, overlapping or neighbouring lines are not accounted for and can introduce spurious effects visible in the continuum of the CCF and impact the shape of the CCF itself. Furthermore, when cross-correlating with a model, the associated uncertainties may be inaccurate as they are not fully propagated through the CCF analysis itself. For HARPS CCFs constructed using DRS (Data Reduction Software) pipeline 3.5 uncertainties are not provided. Uncertainties are therefore estimated from the standard deviation of the flux in the continuum of the CCFs and hence do not arise directly from the data. This is not the case for subsequent DRS releases, which do provide uncertainties propagated from the data.
\newline \indent Least-squares deconvolution (LSD) offers an alternative technique that deconvolves the entire spectrum to produce a single mean-weighted line profile representative of the stellar line profile. LSD directly fits the spectrum and hence produces realistic profiles and profile errors \citep{1997MNRAS.291..658D}. This direct-fit also minimises the contribution of neighbouring and overlapping lines to the final LSD profile. 
\newline \indent In this paper, we introduce $\codename$ (Accurate Continuum fItting and Deconvolution)\footnote[1]{https://github.com/ldolan05/ACID}, a novel technique combining LSD and Markov-Chain Monte-Carlo (MCMC) methods to extract high precision stellar line profiles from high-resolution spectra. The LSD technique and the issues associated with it are introduced in Section \ref{sec:LSD}. In Section \ref{sec:ACID}, we outline the main improvements $\codename$ has over traditional LSD. The implementation of $\codename$ and a demonstration of its capabilities on simulated data are also presented in Section \ref{sec:ACID}. In Section \ref{sec:data}, $\codename$ is used to extract high-precision line profiles from HARPS data for the planet HD189733 and reproduce the Reloaded RM analysis done by \cite{2016A&A...588A.127C}. Our conclusions are presented in Section \ref{sec:conclusion}.

\section{Least-Squares Deconvolution (LSD)}
\label{sec:LSD} 
Least-squares deconvolution (LSD) was first introduced by \cite{1997MNRAS.291..658D} to extract high S/N `mean' line profiles from high resolution \'{e}chelle spectra. Since then it has been successfully used to investigate stellar magnetic fields \citep[e.g.][]{2014MNRAS.444.3517M, 2020MNRAS.493.1130R, 2016AJ....152..207R, 2007A&A...474..969S} as well as extract high resolution RV measurements \citep[e.g.][]{2012MNRAS.424..591B, 2022MNRAS.513.5328L}. 
\newline \indent A stellar spectrum can be modelled by the convolution of a line list, containing the central wavelength and relative absorption depth of each unbroadened spectral line for the normalised spectrum, and a line profile. LSD essentially reverses this process by deconvolving a stellar spectrum, using a given line list, and using the least-squares statistic to obtain a best-fit stellar line profile for that spectrum. In this process the assumption is made that all stellar lines can be modelled by one `mean' line profile and for each line, the `mean' profile is scaled depending on the relative strength of the absorption lines. For the case of overlapping lines, LSD assumes their relative strengths add linearly and deconvolves the overlapping lines accordingly \citep{2010A&A...524A...5K}. The LSD technique and its underlying assumptions are summarised in the following sections.

\subsection{LSD Technique}
\label{sec:LSD_tech} 

In this section we outline the LSD technique as outlined in \cite{collier2001} based on the technique introduced by \cite{1997MNRAS.291..658D}. LSD requires the input of a stellar line list that contains the wavelengths and predicted stellar line depths for a specific wavelength range. These can be obtained from Vienna Atomic Line Database (VALD) \citep{2019ARep...63.1010P, 2015PhyS...90e4005R, 2000BaltA...9..590K, 1999A&AS..138..119K, 1997BaltA...6..244R}. 


A major advantage of LSD is that it takes into account how individual absorption lines blend together to form a feature in the spectrum. Consider a spectrum where $j$ represents the spectral pixel position, and a line list containing lines at wavelengths, $\lambda_i$. For a given spectral wavelength, $\lambda_j$, a contribution is calculated for each absorption line in the line list. This contribution is dependent on the depth of the absorption line, $d_{i}$, and the weight given to that absorption line for a given wavelength pixel in the spectrum, $\Delta x$. A triangular interpolation function is used to spread the weighting of a line over two pixels if the wavelength of the line, $\lambda_i$, does not exactly match the central wavelength of a pixel, $\lambda_j$. This weight, $\Delta x$, can be defined such that:

\begin{equation}
    x=\frac{\nu_k - c\frac{\lambda_j - \lambda_i}{\lambda_i}}{\Delta\nu}
	\label{eq:x}
\end{equation}
\newline $\Delta x = 1 + x$    \indent for \indent  $-1<x<0$
\newline $\Delta x = 1 - x$    \indent for \indent  $0\le x<1$
\newline $\Delta x = 0$        elsewhere.
\vspace{10pt}

\noindent where $v_k$ represents the velocity pixel in the LSD profile and $\Delta v$, is the velocity pixel size. This weight is convolved with the tabulated line depth, $d_{i}$, to calculate the final contribution of the absorption line at position $i$, to the wavelength pixel at position $j$, and is stored in the matrix $\alpha_{jk}$ such that:

\begin{equation}
    \alpha_{jk}=\sum_i{d_{i}\Delta x}
	\label{eq:alpha}
\end{equation}

\noindent The matrix $\alpha_{jk}$ is therefore independent of spectral flux under the assumption that the spectral continuum has been removed and the observed spectrum is therefore normalised. In this case, $\alpha_{jk}$ depends only on the spectral wavelengths and predicted stellar lines. The deconvolved least-squares profile, $z_k$, can then be obtained from the normalised observed spectrum, $r_j$, by minimising:

\begin{equation}
    \chi^2 = \sum_j{ \left( \frac{(r_j-1) - \sum_k{\alpha_{jk}z_k}}{\sigma_j} \right)}^2
	\label{eq:chi2}
\end{equation}

\noindent The solution for the least-squares profile and the uncertainties, $\sigma$, can be written in matrix form considering $S$ as the diagonal matrix containing $\frac{1}{\sigma_j}$ and $R$, the matrix containing $(r_j-1)$. Here, the superscript $T$ denotes the transpose of a matrix.

\begin{equation}
    z = (\alpha^T\cdot S^2 \cdot \alpha)^{-1} \alpha^T\cdot S^2\cdot R
	\label{eq:matrix_solution}
\end{equation}

\begin{equation}
    \sigma_{z}^2 = Diag(\alpha^T\cdot S^2\cdot \alpha)^{-1}
	\label{eq:prof_errors}
\end{equation}

\label{sec:new_od} 
\begin{figure*}
\includegraphics[width=\textwidth, trim = {0, .7cm, 0, .7cm}, clip]{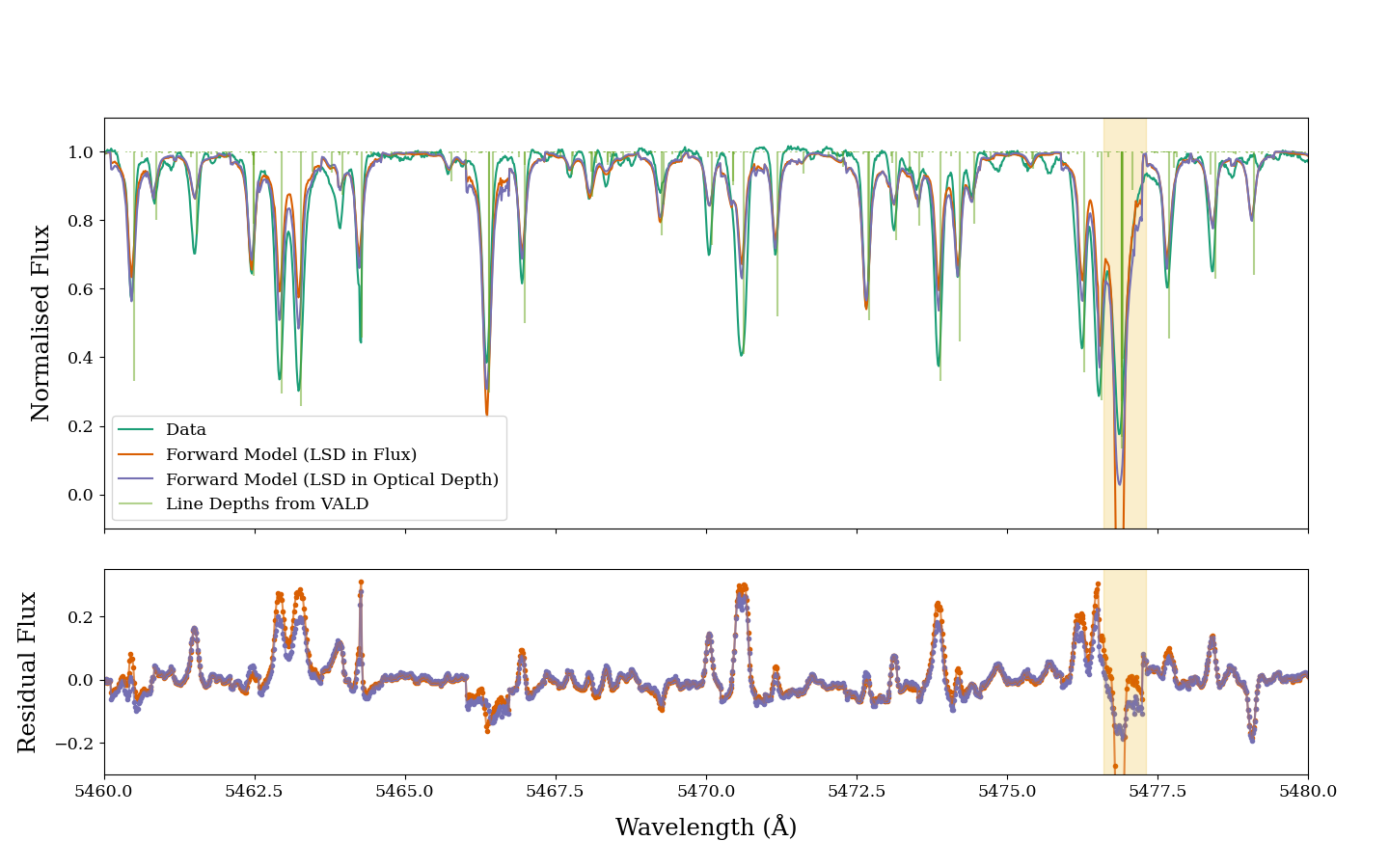}
    \caption{The two forward models for LSD performed in flux and effective optical depth are plotted on top of the observed data in the top panel. One example of a deep absorption feature is highlighted. It can be seen for this feature that LSD performed in flux predicts a nonphysical normalised line depth of $\sim2.5$. This depth is greatly improved when LSD is performed in effective optical depth. For reference, the line depths provided by the VALD line list are also plotted. The residuals for each of these models are presented in the bottom panel. It can be seen that the line depths for deep, overlapping lines are in better agreement when LSD is performed in effective optical depth compared to LSD performed in flux.}
    \label{fig:flux_od_LSD}
\end{figure*}

\noindent It is important to note that the observed spectrum must be continuum normalised before LSD can be performed. This is because the line depths given in the line list are normalised, but it is also necessary to remove any structure present in the continuum that could affect the shape of the deconvolved line profile. This application of LSD relies on a number of assumptions that, if ignored, can affect how the resulting LSD profiles are interpreted. We now discuss some of the issues with the conventional application of LSD and outline how we can circumvent some of these.

\subsection{Issues with LSD}
\label{sec:LSD_issues}

\subsubsection{Linear Addition of Overlapping Lines}
\label{sec:LSD_od_issue}



\indent A core assumption of traditional LSD is that the relative strength of overlapping lines adds linearly. As LSD is carried out in flux, this leads to the assumption that overlapping intensities add linearly and contribute to LSD's potential to greatly overestimate the depth of stellar lines, specifically in the case of strong blends, i.e. very deep lines or regions with many overlapping lines. This is acknowledged in \cite{1997MNRAS.291..658D} and they state that they ensure that the sum of normalised depths never exceeds 1. However, they do not provide any further detail or outline reproducible steps on how this can be done using traditional LSD. Furthermore, this effect comes from an underlying assumption in traditional LSD that affects how LSD treats overlapping lines. Therefore, arbitrarily correcting the line depths to avoid this scenario would still result in discrepancies between the LSD forward model and observed data. Fig.~ \ref{fig:flux_od_LSD} shows examples of model spectra based on LSD profiles derived by either performing LSD in flux or in effective optical depth. In this case, LSD was applied to sample data (outlined further in Sections ~\ref{sec:LSD_od} and ~\ref{sec:data}) and the resulting profile was used to create a normalised forward model spectrum, $f_j$, via:

\begin{equation}
    f_j=\sum_k{\alpha_{jk}z_k} + 1
	\label{eq:predicted_spec}
\end{equation}

The $\alpha_{jk}$ used is constructed using a line list extracted from VALD using the `extract stellar' query with a detection threshold, microturbulence, effective temperature and $\log{g_{*}}$ of 0.001, $1$\,km\,s$^{-1}$, 5022\,K \citep{2011A&A...530A.138C} and 4.56 \citep{2015MNRAS.447..846B}, respectively. These parameters were chosen to best match the parameters of HD189733 (which we later used to test our method in Section \ref{sec:data}). VALD returns a line list with the closest input parameters for effective temperature, $T_{eff}$ and $\log{g_{*}}$ to those provided. In this case, a line list for $T_{eff}$\,=\,5000\,K and  $\log{g_{*}}$ = 4.5 was returned. It can be seen that when LSD is performed in flux the depths of deep absorption features are generally overestimated. For example, the strong blend at $\sim5475.4$\,\AA~ leads to a prediction of a much deeper feature with a nonphysical normalised depth of $\sim2.5$.
\newline\indent This issue with the traditional approach to LSD was investigated in detail by \cite{2010A&A...524A...5K}. In the case of blending two identical Fe I lines together it was found that as the profile depth of the lines increased so did the discrepancy between the linearly added profiles. They concluded that the assumptions of line profile self-similarity and the linear addition of blends break down for lines deeper than 20$\%$ of the continuum intensity. \cite{2010A&A...524A...5K} attempted to remedy this by invoking a multi-line LSD technique. This assumes that the stellar spectrum consists of $n$ different mean line profiles and hence can assign different profiles (and profile depths) to weak and strong lines. This promises to address the issue of over-estimated line depths \citep{2010A&A...524A...5K}. However, the S/N of a resulting LSD profile is dependent on the number of lines used to construct that profile and hence producing separate profiles for shallow and deep lines would therefore spread the S/N gain across the profiles. An alternative method is therefore preferable that can resolve this issue while maximising the S/N when applied to observational data. 

\subsubsection{Stellar Continuum Correction}
\label{sec:Stellar Continuum Correction}

\begin{figure}
	\includegraphics[width=\columnwidth]{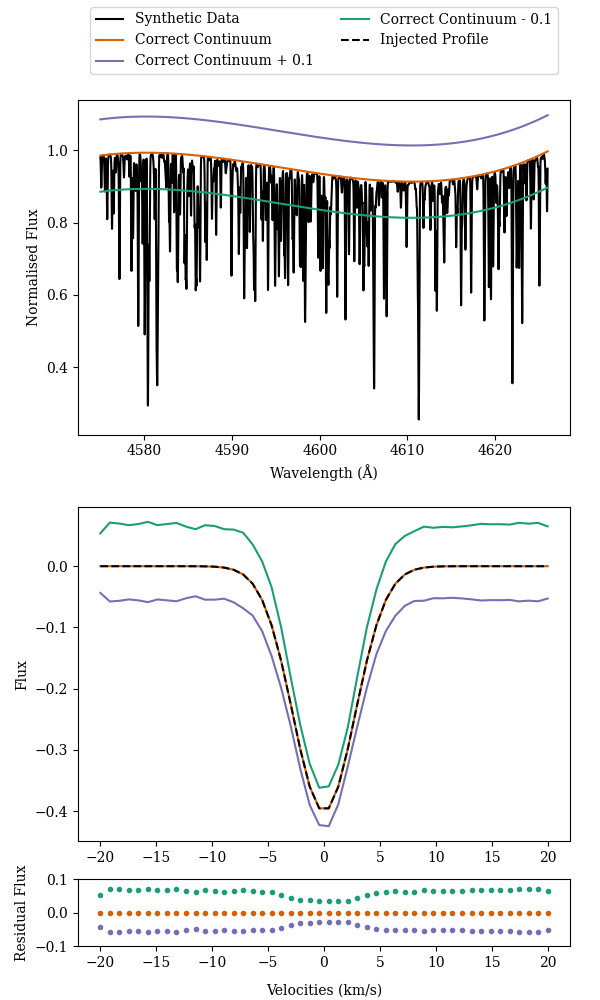}
    \caption{The upper panel shows the model spectrum (created from the convolution of the injected profile and a stellar line list) with each of the applied continuum corrections (vertically offset by $\pm0.1$). The middle panel shows the LSD profiles that result from the spectra corrected by each of the applied continuum corrections and the injected profile (dashed line). The residuals are shown in the bottom panel.}
    \label{fig:LSD_continuum}
\end{figure}

Another assumption of LSD is that the spectrum must be normalised before LSD is applied. This therefore assumes that the stellar continuum has been fully removed with the consequence that an inaccurate continuum correction can cause distortions in the resulting LSD profiles. Typically, the continuum is found and corrected by fitting the top of the observed spectrum, i.e. the pseudo continuum, and dividing the spectrum by this fit. This technique, however, can be very sensitive to noise as well as spectral regions containing a high density of weak absorption lines. High noise levels can, for example, effectively raise the pseudo-continuum as the top of the noise is fit, which lies above the real continuum level. In the case of regions containing a high density of absorption lines, this results in very little to no wavelength range free of absorption features. Within this range, the overlapping lines blend together and act to lower the pseudo-continuum. This lowering/raising of the pseudo-continuum can therefore result in contamination from the continuum remaining in the spectrum.
\newline \indent Simulated data was used to investigate the extent of an inaccurate continuum correction on the resulting LSD profile. A model normalised spectrum was created through the convolution of a line list with an injected profile. The model spectrum was constructed on a wavelength grid of pixel size 0.01\,\AA~across a wavelength range equivalent to a single order in a HARPS spectrum (4575-4626\,\AA). The stellar line list used was the same as outlined in Section \ref{sec:LSD_od_issue}. A Gaussian of normalised depth of 0.4 and a FWHM $\sim 6.6$ \,km\,s$^{-1}$ was used as the injected profile. The forward model from the injected profile and line data was multiplied by a third-order polynomial to act as the model continuum. The exact polynomial used here is irrelevant as it serves to demonstrate how residual continuum left in a spectrum can impact the resultant LSD profiles. LSD was performed on the continuum-corrected model data and the resulting profiles were compared to the injected profile. The same line list was used in the creation of the model data as was used for the application of LSD ensuring that any effect observed is solely from the continuum correction applied. The model data was continuum corrected by dividing the spectrum by a selected continuum fit. 
\newline \indent Fig.~\ref{fig:LSD_continuum} shows the results for the case where the continuum of the normalised profile is vertically offset by 0.1, for a positive and negative case. Our results show that when the correct continuum is used, i.e. LSD is applied to a perfectly normalised spectrum, a noiseless profile is returned that matches the injected profile. This is expected as LSD is the exact reversal of the convolution used to create the model data. However, correcting with a slightly offset continuum fit was found to change the depth and shape of the resulting profile as well as induce spurious noise that is visible in the profile's continuum. As mentioned, the normalised depth of the injected profile was set to 0.4. When the true continuum correction was used the normalised profile depth was retrieved to within $1\times10^{-10}$ accuracy and the resulting profile contained very little noise with an out-of-line RMS of 0.0001 in normalised flux. In the case of a continuum correction exhibiting a $\pm0.1$ offset from the true continuum position, the normalised profile depths are seen to change with the measured depths being $0.392\pm0.002$ and $0.408\pm0.002$ for a positive and negative offset, respectively. These depths were measured by a Gaussian fit to the final profiles where the uncertainties given account for the effect of the increased noise in these profiles. The out-of-line RMS also increases for a positive and negative offset to 0.004 and 0.005, respectively. No noise was added to the model data and so an incorrect continuum correction can be seen to increase the out-of-line RMS by a factor of 40-50 times that seen from a correct continuum correction. This highlights how important a correct continuum fit is to the reliability of the final LSD profile.

\section{\codename}
\label{sec:ACID} 
Our novel technique: $\codename$ aims to remedy these issues and returns accurate line profiles from high-resolution spectra. $\codename$ builds on current LSD techniques by performing LSD in effective optical depth (outlined in Section \ref{sec:LSD_od}) and simultaneously fitting the continuum and the mean line profile of each spectrum (outlined in Section \ref{sec:LSD_cont}). 
$\codename$ requires the input of the spectral data (including wavelength, flux, errors and average S/N of each spectrum) as well as a theoretical line list. Currently, $\codename$ accepts line lists in the short- and long-form format provided by VALD.
VALD provides a spectral line list based on the stellar parameters given by the user such as the stellar temperature, log$\,g$, microturbulence velocity, chemical composition (including metallicity), the minimum detection threshold (i.e. minimum line depth of lines included in the line list) and a specified wavelength range. 

\subsection{Main Improvements from Traditional LSD}
\label{sec:LSD_improvements}

\subsubsection{LSD in Effective Optical Depth}
\label{sec:LSD_od}
 $\codename$ addresses the overestimation of the depths of overlapping lines that have been previously shown with traditional LSD. This issue arises from working in flux as the assumption that overlapping lines add linearly breaks down when dealing with overlapping intensities for deep or very dense regions of absorption lines. $\codename$ solves this problem by working in effective optical depth where the assumption of linear addition still holds. For this reason, all LSD processes in $\codename$, including convolution, are performed in effective optical depth, $\tau$. This can be calculated for a spectral wavelength pixel, $j$, from the normalised flux, $F_j$ by:

\begin{equation}
    \tau_{j}=\ln({F_j}) + 1
	\label{eq:flux2od}
\end{equation}

\begin{figure*}
\includegraphics[width=\textwidth, trim = {0, .7cm, 0, .7cm}, clip]{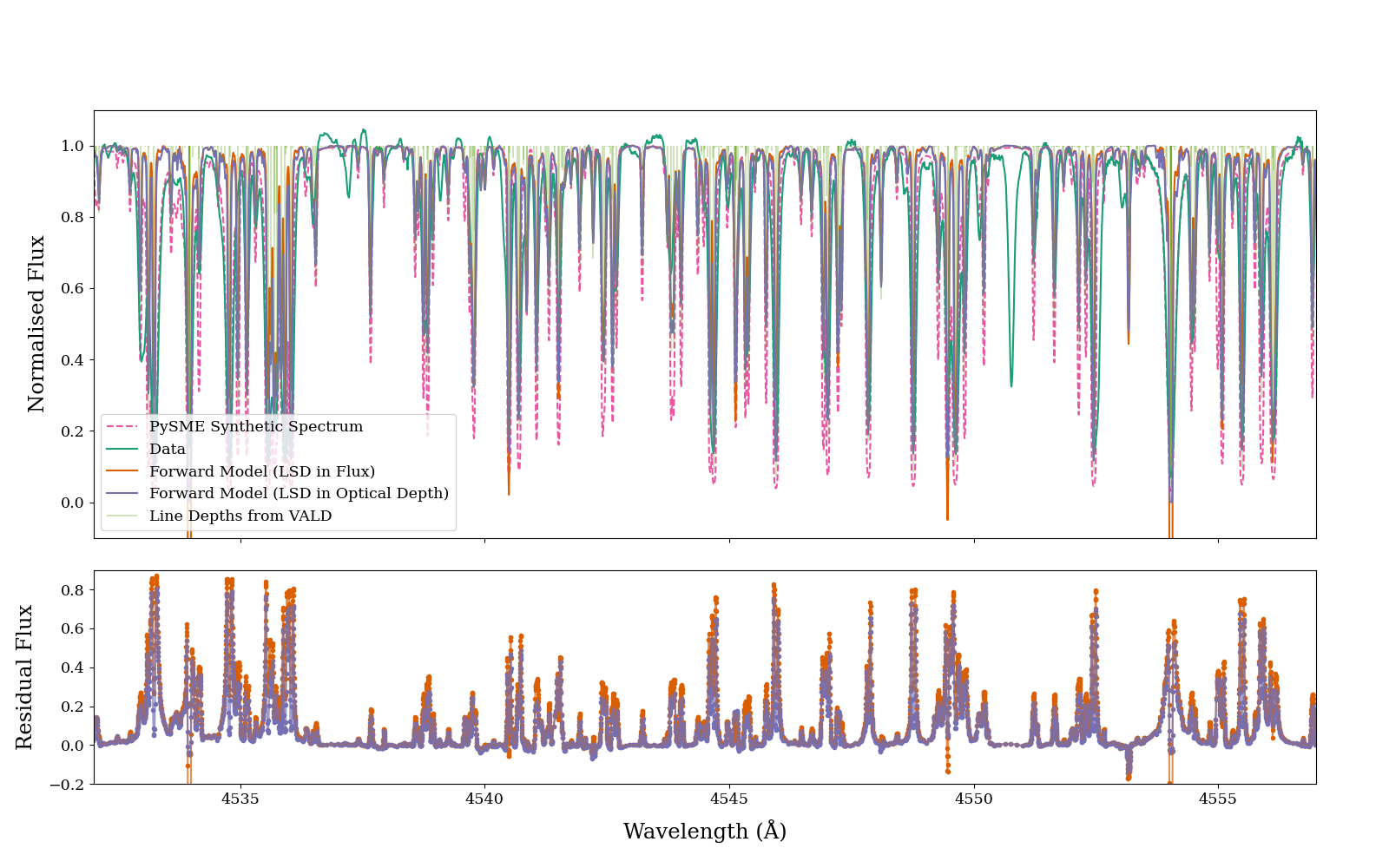}
    \caption{PySME was used to construct a synthetic spectrum (plotted in the top panel). LSD was then performed on this synthetic spectrum in both flux and effective optical depth. The two forward models for LSD performed in flux and effective optical depth are also plotted. For reference, the observational data (from Figure \ref{fig:flux_od_LSD}) is plotted in the top panel as well as the line depths provided by VALD that was used for the application of LSD. We highlight the lines located at $\sim4537$\,\AA~ and $\sim4551$\,\AA~ visible in the observational data that are not predicted in either the LSD models or synthetic spectrum.}
    \label{fig:SME}
\end{figure*}

\noindent The positive line depths in flux, $d_i(F)$, as provided from VALD, must also be converted into effective optical depth such that:
\begin{equation}
    d_{i}(\tau) = -\ln({1-d_i(F))}
	\label{eq:depths2od}
\end{equation}
\noindent After these conversions, LSD can be applied with the same steps described in Section \ref{sec:LSD_tech} to produce a mean line profile, $z$, in effective optical depth. The resulting profiles and any forward models used are also converted back into flux to produce the final $\codename$ profiles and for comparisons to the observed spectrum.
\newline \indent To test this process, LSD was performed in both flux and effective optical depth on an example spectrum from observations of HD189733 covering a wavelength range of $5431.7 - 5494.1$\,\AA~ (further details on data given in Section \ref{sec:HD189733_data}) and the resulting forward model spectra were compared. The stellar line list used was the same as outlined in Section \ref{sec:LSD_od_issue} with the minimum detection threshold set to 0.001 in normalised flux depth. The same line list and hence number of lines were used for both LSD applied in flux and effective optical depth, amounting to 2310 lines over the wavelength range considered. The forward models produced indicate a better fit to the data when LSD is performed in effective optical depth as significantly fewer overestimated lines are present. Fig.~\ref{fig:flux_od_LSD} shows an example of a small wavelength range so that the behaviour of individual lines can be observed. It can be seen for the strong blend located at $\sim5477$\,\AA, which produces a non-physical depth of $\sim2.5$ when LSD is applied in flux, has a depth in better agreement with the data when LSD is performed in effective optical depth.
\newline \indent We have demonstrated that performing LSD in effective optical depth produces more realistic line depths in the case where there are overlapping lines. However, we acknowledge that other types of line blending that were correctly treated in traditional LSD, such as that induced from rotational broadening, will not be treated correctly under this approach. In this work we are interested in slow rotating stars where this approximation holds, however, in the case of rapid rotating stars where rotational broadening would dominate this approximation would break down and this approach would no longer be applicable.
\newline \indent It can also be noticed that, despite better agreement with strongly blended lines when LSD is performed in effective optical depth, there are still discrepancies in the line depths present in the data compared to the forward model. These discrepancies could arise from either inaccuracies in the stellar line list used to construct the $\alpha_{jk}$ matrix or from the remaining assumptions in the LSD model. To further investigate this we applied a similar test to a detailed synthetic spectrum constructed using \textit{PySME} \citep{2023A&A...671A.171W, 1996SME, 2017SME}. \textit{PySME} produces realistic stellar spectra using radiative transfer calculations from input stellar parameters and a VALD stellar line list. The synthetic spectrum was constructed on the wavelength grid of the HARPS order covering the wavelength region $4506.8 - 4558.4$\,\AA~. This region was chosen as it was seen to contain an obvious discrepancy at $\sim4552$\AA~ between the LSD forward model and the observed data. As before, we used VALD to obtain a line list based on the stellar properties that are closest to those of HD189733, although it should be noted that PySME only takes the line positions from this and not the line depths. To keep the stellar inputs consistent with those used to construct the line list the microturbulence, effective temperature and $\log{g_{*}}$ were set to $1$\,km\,s$^{-1}$, 5000\,K and 4.5, respectively. The macroturbulence and $v\sin i_{*}$ were set to $2$\,km\,s$^{-1}$ and $3.3$\,km\,s$^{-1}$, respectively, and the MARCS stellar models were used \citep{atom_SME} for the stated parameters. LSD was then performed on this synthetic spectrum in both effective optical depth and flux and the respective forward models can be seen in Fig.~\ref{fig:SME}.

\indent It can be seen that, compared to LSD performed in flux, LSD in effective optical depth produces a better fit to the synthetic spectrum. While the greatest improvement is seen in the case of very deep absorption lines we also note that for the case of shallower absorption lines LSD in effective optical depth still produces smaller residuals than traditional LSD. This further proves that LSD in effective optical depth outperforms traditional LSD for slow-rotating stars. As was observed with the application to observational data, discrepancies can be seen between the LSD forward model and the synthetic spectrum. It can also be seen that the depths of many of the absorption features produced from the synthetic spectrum do not match those seen in the observational data (Fig.~\ref{fig:SME}), with larger discrepancies being observed between the synthetic spectrum and observational data than is seen between the observational data and the LSD forward models, despite the same stellar parameters being used. SME does not take an estimate for the line depths from the provided VALD line list but instead takes the line positions from the provided line list. The relative line depths are then calculated from these along with the input stellar parameters using the radiative transfer equations. For this reason, we cannot state if the discrepancies observed between the synthetic spectrum and both LSD forward models arise from inaccurate line depths in the VALD line list or differences in the radiative transfer and LSD models. We can, however, observe that there are clear inaccuracies in the line data that appear in both the LSD forward models and the synthetic model from \textit{PySME}. Examples include the lines located at $\sim4537$\,\AA~ and $\sim4551$\,\AA~ visible in the observational data that are not predicted in either the LSD models or synthetic spectrum (Fig.~\ref{fig:SME}). While this shows evidence that inaccurate line data is contributing to the discrepancies between the observational data and the LSD forward models we cannot confidently state that this is the sole contributor as it could be a combination of this and the underlying assumptions in LSD. Despite this, we have shown that $\codename$ does improve on traditional LSD for both the application to observational data and a synthetic spectrum by performing LSD in effective optical depth.

\subsubsection{Simultaneous Continuum Fit}
\label{sec:LSD_cont}
As well as working in effective optical depth $\codename$ also improves upon traditional LSD by simultaneously fitting the line profile and the stellar continuum. Section \ref{sec:Stellar Continuum Correction} highlighted that an incorrect continuum correction directly affects the shape of the profile that LSD produces and results in additional noise. The simultaneous fit of the continuum and LSD profile provides a better constraint on the continuum as the noise in the stellar line profile will be minimised when the continuum fit has converged correctly. This simultaneous fit works by comparing the forward model directly to the data. For a given profile, $z$, a model spectrum (including the continuum), $m_j$, can be constructed by the multiplication of the forward model spectrum, $f_j$, and the model continuum fit (modelled by an $n^{th}$ order polynomial) such that:
\begin{equation}
    m_j = C(c_0 + c_1\lambda_{j} + c_2\lambda_{j}^2... + c_n\lambda_{j}^{n}) * f_j
	\label{eq:model}
\end{equation}
\noindent where $C$ is a scaling factor and $c_n$ are the polynomial coefficients describing the shape of the continuum. Since $\codename$ performs LSD in effective optical depth, the forward model spectrum is first converted from effective optical depth into flux before the continuum fit is added. The resulting model spectrum can then be directly fit to the data to obtain both a best-fit line profile and continuum fit. This is done using MCMC methods and will be further discussed in the following sections.

\subsection{Implementation of \codename}
\label{sec:Implementation}



 \subsubsection{Combining Spectral Frames}
 The spectral continua, including any telluric and systematic effects, are expected to be consistent over an observation night. This allows a master continuum fit to be applied to a high S/N combined spectrum.  $\codename$ fits the continuum for all frames on an order-by-order basis, applying the same correction to each individual frame. The frames are first adjusted to sit on the same continuum level by dividing each spectral frame by a reference frame (taken as the highest S/N frame for that observation night). A 4\textsuperscript{th}-order polynomial is fit to this ratio, after removing outliers, and used to adjust each spectrum. These adjusted spectra are retained and will be corrected by the final continuum fit for a given order (outlined later in this section). 
 \newline \indent A combined spectrum is constructed using a weighted mean of the adjusted frames with weights that are calculated based on the average S/N of each spectrum. A S/N for the combined spectrum is also calculated using a weighted mean. Not only does fitting the continuum to a combined spectrum obtain a more accurate fit due to the higher S/N but it also maintains consistency between the frames and ensures that the profile depth remains consistent. If only one spectral frame was available for each observation night, i.e. for the case where there are many singular observations scattered over different observation nights, then the continuum fit would still need to be applied on a night-by-night basis. The combined spectrum is then taken as the single frame for the night. The following steps are applied in the same way with only a single LSD profile being produced.
 
\subsubsection{Masking}
\label{sec:masking}

\indent It is necessary to exclude regions of the stellar spectrum that could affect the mean stellar line profile such as those affected by strong telluric contamination and very wide absorption lines that diverge from the shape of the majority of stellar lines. To remedy this, three stages of masking are performed on the combined spectrum. Masking is kept consistent between the frames for a given observation night to ensure any frame-dependent line profile distortion is from an astrophysical source. In the first two masking stages problematic areas are identified by calculating residuals between the combined spectrum and a forward model spectrum constructed from the initial continuum fit and LSD line profile. An initial continuum fit is found using a window function that isolates the pseudo-continuum of the spectrum by fitting a polynomial using a window function. The window function isolates the maximum flux in 100-pixel bins, which acts as the pseudo-continuum. The bin size can be adjusted by the user, however, the default size of 100 pixels was seen to cover multiple lines in the spectra for HD189733. This ensured that the maximum flux value came from a pseudo-continuum value instead of the central part of an absorption feature. The polynomial is then fit to this pseudo-continuum. LSD is then performed on the initial continuum corrected spectrum and the returned profile is taken as the initial profile.
\newline \indent The first masking stage removes regions affected by strong telluric contamination and known broad lines that are expected to deviate from the average line shape. Although many of these lines will be masked as a result of additional steps outlined later, as an extra precaution wavelength regions known to contain these lines are manually masked out first to ensure they do not affect the final profiles. By default, the regions covering $6280-6350$\,\AA~ and $6800-7900$\,\AA~ are masked as these include strong telluric regions such as the oxygen A and B bands. The user can then specify other wide lines to be masked such as the Balmer lines, Calcium H and K lines and Sodium doublet. In the case of user-specified lines $\sim3.5$~\AA~ is masked out on either side of the line.
\newline \indent In the second stage, broad absorption lines that diverge from the average profile shape are removed. In this stage, areas with more than 20 consecutive pixels in which the spectrum and forward model deviate by more than 25\% of the spectral flux value are masked. This removes large regions that deviate from the forward model and absorption features that could affect the mean line profile including very wide absorption lines (such as the Balmer lines, Ca H and K lines and the Na doublet) and extremely dense regions of lines. 
\newline \indent The third stage of masking is applied to remove individual lines that vary greatly from the predicted depths. This is done by masking pixels where the residuals between the spectrum and the forward model deviate more than $n\sigma$ from the median of the residuals, where $\sigma$ denotes the standard deviation and n is an integer number ($n=1$ being the default). This step is necessary as discrepancies in line depths can affect the continuum fit when using the MCMC method. For example, if many lines were deeper in the data compared to what is predicted by the line list then this will act to artificially pull the continuum fit down to compensate. This stage of masking is only applied during the MCMC continuum fit to the combined spectrum and is removed before the final run of LSD (Section \ref{sec:final_prof}) to obtain the final $\codename$ profiles. At all stages, the number of consecutive pixels, residual deviation percentage, $n\sigma$ cut-off and telluric regions can be tailored by the user.

\subsubsection{MCMC Approach for Continuum Fitting}
\label{sec:MCMC_cont}
 After masking has been applied the combined spectrum is used to find the continuum fit using an MCMC approach. The python package \texttt{emcee} \citep{2013PASP..125..306F} is used to simultaneously fit the mean line profile, continuum scaling factor and polynomial coefficients by comparing the model spectrum (equation \ref{eq:model}) directly to the combined observed spectrum. The continuum scaling factor, polynomial coefficients and every single point in the mean (LSD) line profile are set as free parameters. This ensures that the depth, shape and continuum level of the mean line profile are not biased in any way. 
\newline \indent Throughout this work, a 3\textsuperscript{rd}-order polynomial was used for the continuum fit as this allowed the continuum to be well constrained for our data sets without the added computation time associated with higher orders. The order of polynomial to be applied can also be adjusted by the user. The initial continuum coefficients (continuum scaling factor and polynomial coefficients) are taken from a fit performed using a window function that takes the maximum point from every unmasked 10 pixels from the combined spectrum. The input profile comes from an initial run of LSD on the continuum-corrected spectrum (corrected using the input continuum coefficients). The resulting $\alpha_{jk}$ matrix (containing the line data) from this is saved as a global array. This saves computation time as the $\alpha_{jk}$ matrix does not need to be recalculated during each iteration of the MCMC fit.
\newline \indent In general, the minimum detection threshold applied to the line list (i.e. the minimum depth of absorption line that would be considered) was also found to affect the continuum fit of the spectrum. Specifically, for the case when the minimum detection threshold is set higher than the noise level present in the spectrum the resulting continuum fit is higher than expected resulting in a downward offset in the resulting LSD profile. This occurs as very small absorption lines that are detectable (i.e. have a line depth deeper than the noise level) in the data are removed from the line list and treated as part of the continuum. The obvious way to remedy this is to set the minimum detection threshold to 0, thereby not excluding any lines. This, however, is very computationally intensive and can introduce noise as many undetectable lines are contributing to the final profile. To remedy this a balance must be struck between including necessary lines and removing those that only add to the computational load. In $\codename$, the minimum detection threshold is based on the average S/N of the spectrum such that our detection threshold is equal to $\frac{1}{3(S/N)}$. This ensures that all detectable lines are included in the line list while still minimising the computational load. VALD generally provides the stellar line depth to 3 significant figures which, in the case of high-resolution spectra, would lead to the calculated detection threshold to be smaller than the accuracy provided by VALD. In this case, we are limited by the line list, which could lead to very weak lines being excluded.
 \begin{figure*}
	\includegraphics[width=\textwidth, trim={0cm, .8cm, 0cm, .8cm}, clip]{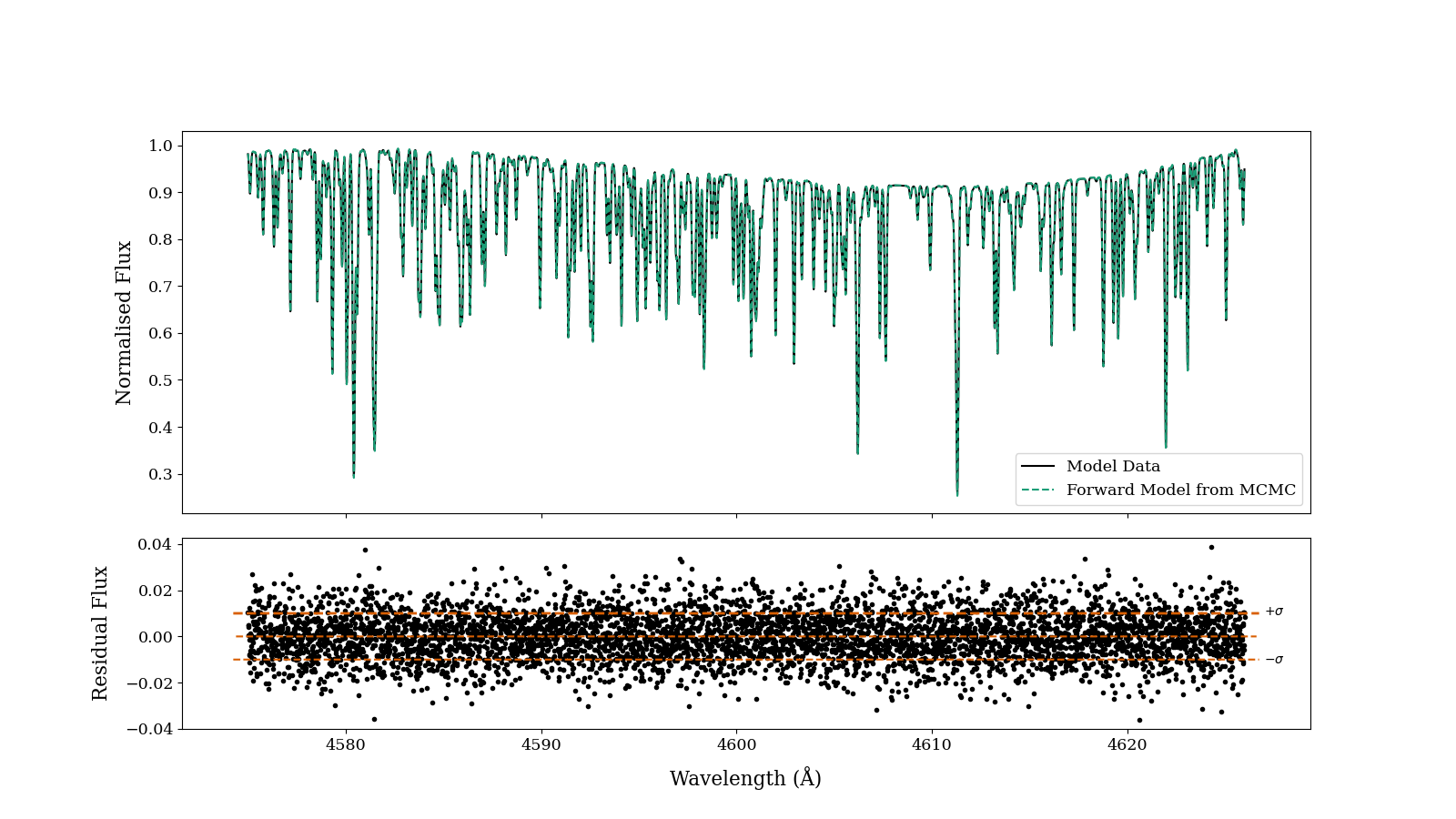}
    \caption{The model spectrum used to test $\codename$ is shown in the top panel with the best fit forward model from MCMC plotted on top. The residuals are shown in the bottom panel with the -1$\sigma$, 0 and 1$\sigma$ limits indicated by dashed lines. It can be seen that $\codename$ accurately returns the continuum fit and model line profile.}
    \label{fig:mcmc_syn}
\end{figure*}

\begin{figure}
	\includegraphics[width=\columnwidth, trim={0cm, 0.65cm, 1.1cm, 2cm}, clip]{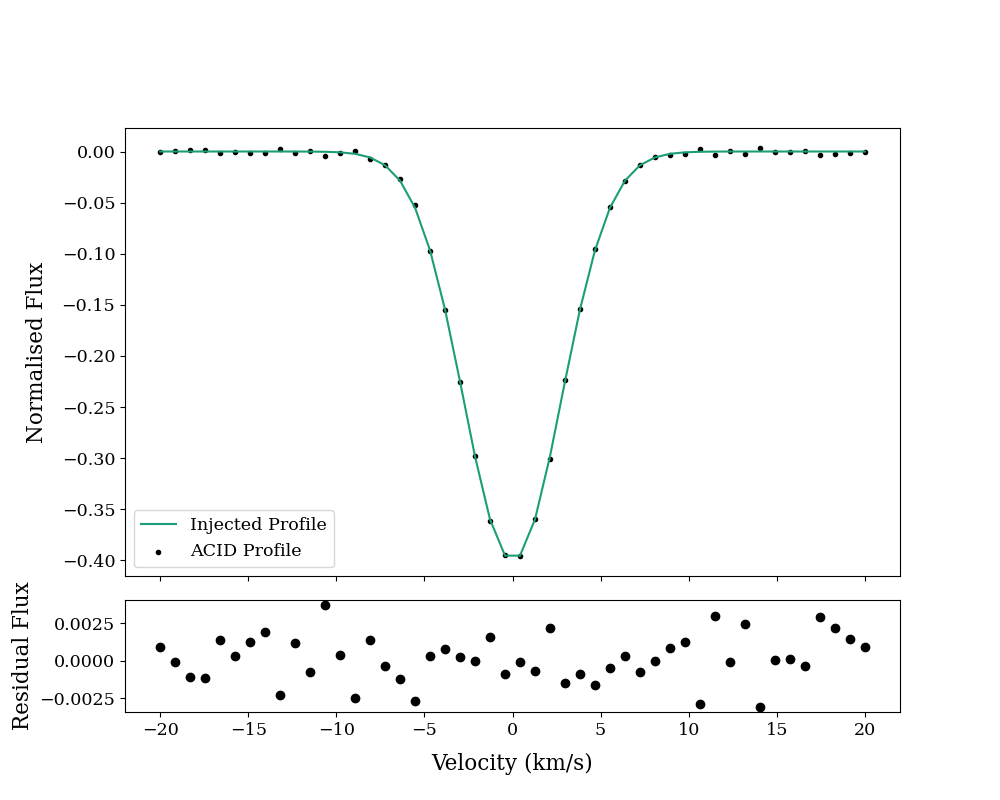}
    \caption{The top panel shows the resulting $\codename$ profile obtained from the model data shown in Fig.~\ref{fig:mcmc_syn}. The injected profile used to construct the model spectrum is also plotted for comparison with the residuals shown in the bottom panel.}
    \label{fig:mcmc_syn_prof}
\end{figure}
\subsubsection {Obtaining the Final Profiles}
\label{sec:final_prof}
After the best-fit continuum has been found the individual frames that were adjusted to sit at the same continuum level are used to obtain the final mean line profiles. Each frame is continuum-corrected using the output polynomial from the MCMC fit. The mask outlined in Section \ref{sec:masking} is interpolated onto the wavelength grid for the given frame. LSD is performed on each masked and continuum-corrected spectrum to retrieve the final profile and associated errors. This final run of LSD ensures the associated errors arise from the data and not the quality of fit from the MCMC. The errors from the continuum fit are propagated onto the continuum corrected spectrum so that the final LSD errors account for the uncertainty in the continuum fit also. $\codename$ allows the user to input the velocity grid for the final profiles. In our analysis, the velocity grid was chosen so that the resulting profile was centred on the grid. In all cases, the velocity pixel size was calculated based on the average wavelength separation in the spectrum ($\sim$0.82~km\,s$^{-1}$ for HARPS). $\codename$ repeats the steps listed above for each order so that a profile is obtained for each order and frame combination. 
\newline \indent A combined profile and the associated errors were then calculated for each frame using a weighted mean of the subsequent order profiles. The final profile was interpolated onto a velocity grid consistent across all frames, again with a constant velocity pixel size equal to that of the original velocity grids. This allowed for profile comparisons to be easily made between frames. 

\subsection{Application to Model Data}
\label{sec:syndat}

\indent $\codename$ was applied to a model spectrum constructed via equation \ref{eq:model}. The injected profile was modelled as a simple Gaussian with a normalised depth of 0.4 and FWHM $\sim 6.6$~km\,s$^{-1}$). The model continuum was described by a 3\textsuperscript{rd}-order polynomial. The $\alpha_{jk}$ matrix was constructed using line data obtained from VALD using the same stellar line list outlined in Section \ref{sec:LSD_od_issue}. Synthetic noise was added using a Gaussian distribution with a standard deviation of 0.01. This produced a similar noise level to that seen in HARPS data for HD189733 -- outlined in Section \ref{sec:HD189733_data}. The error on the model spectrum was set to 0.01 for all pixels. 
\newline \indent A comparison of the model spectrum and the best fit model from \texttt{emcee} can be seen in Fig.~\ref{fig:mcmc_syn}. $\codename$ retrieves the `true' continuum fit with the forward model of the continuum fit and profile providing good agreement with the spectrum. This results in the retrieval of the injected profile within the induced noise levels (shown in Fig.~\ref{fig:mcmc_syn_prof}), demonstrating $\codename$'s ability to find and correct the true continuum of the spectrum and return the true line profile to a high precision.

\section{Application to HARPS Data}
\label{sec:data}
\subsection{Data}
\label{sec:HD189733_data} 
This work used the observational data of HD189733 from the HARPS \'echelle spectrograph on the ESO 3.6 m telescope in La Silla, Chile. Four observation nights, obtained during the transit of HD189733b, were used: 2006 July 29/30, 2006 September 7/8, 2007 July 19/20 and 2007 August 28/29. These particular data sets have been studied extensively in the literature (\citealt{2009A&A...506..377T}; \citealt{2010MNRAS.403..151C}; \citealt{2015A&A...577A..62W}; and \citealt{2016A&A...588A.127C}). Full transits of HD189733b are covered in all of the observation nights except for the night during July 2006. Here, only a partial transit is observed due to poor weather conditions. In total, there are 111 frames, taken both in and out of transit. 
\newline \indent Throughout this work the HARPS e2ds spectra produced from the DRS pipeline version 3.5 were used. These are 2-dimensional \'echelle order spectra. Before spectra were provided to $\codename$ they were blaze corrected using the blaze files identified from the FITS header of the provided e2ds file. The spectra were also corrected to remove the Barycentric Earth Radial Velocity (BERV) from each frame using the BERV correction provided in the FITS header of the e2ds file. ~$\codename$ is also compatible with the 1-dimensional combined spectra (s1d) from the DRS pipeline version 3.5 although, due to the oversampling of this onto a finer wavelength grid and the merging of the orders into one spectrum, we opted to use the e2ds spectra throughout. For the case of s1d spectra, $\codename$ splits each spectral frame into smaller wavelength ranges corresponding to those of the e2ds orders. To avoid overlap regions being processed twice, the overlap region occurring at longer wavelengths is discarded for each wavelength range. $\codename$ then applies the same method outlined for the e2ds spectra where these wavelength ranges act as the orders in the e2ds case. As the continuum is then fit to these smaller wavelength ranges, we avoid the need for high-order polynomials or a more complicated continuum model to describe the continuum of the entire s1d spectrum.
\newline \indent For comparison to the resulting $\codename$ profiles, we use the corresponding CCFs provided from the HARPS pipeline. For these observations, the CCFs available on the archive are processed using different stellar masks for different nights. Namely, a K5 stellar mask was applied for observations taken in July and September 2006 and a G2 stellar mask was applied for July and August 2007. We requested for all 4 observation nights to be reprocessed using a K5 stellar mask which better matched the spectral type of HD189733 (K2). These reprocessed CCFs are used throughout this work.
\newline \indent For the observations taken on July 2007 a feature was observed in the resulting $\codename$ profiles (further discussed in Appendix \ref{sec:feature_sec}). This was found to only appear in orders 40-42 and is not present for any other observation night nor in the corresponding CCF profiles. For this reason, orders 40-42 were excluded for July 2007 only. In the application of $\codename$ to HD189733, we mask out lines at positions 3820.33, 3933.66, 3968.47, 4327.74, 4307.90, 4383.55, 4861.34, 5183.62, 5270.39, 5889.95, 5895.92\AA, some of these regions being the Balmer lines, Calcium H and K lines and Sodium doublet lines. These are masked out in the first masking stage outlined in Section \ref{sec:masking}.


\begin{figure*}
        \includegraphics[scale = 0.6, trim={0, 0, 1.2cm, 1.3cm}, clip]{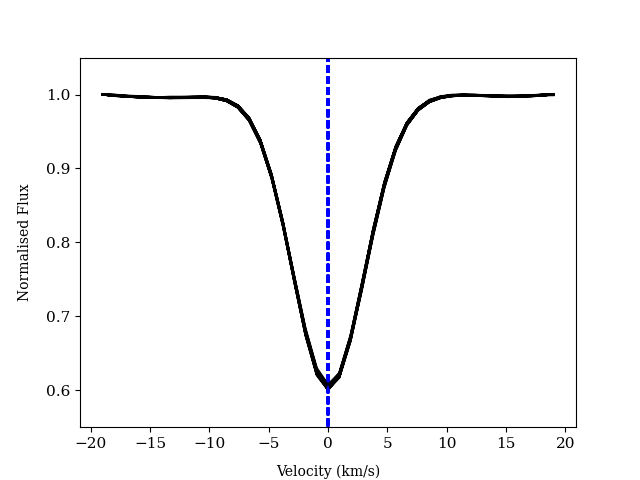}
	\includegraphics[scale = 0.6, trim={2cm, 0, 0, 1.3cm}, clip]{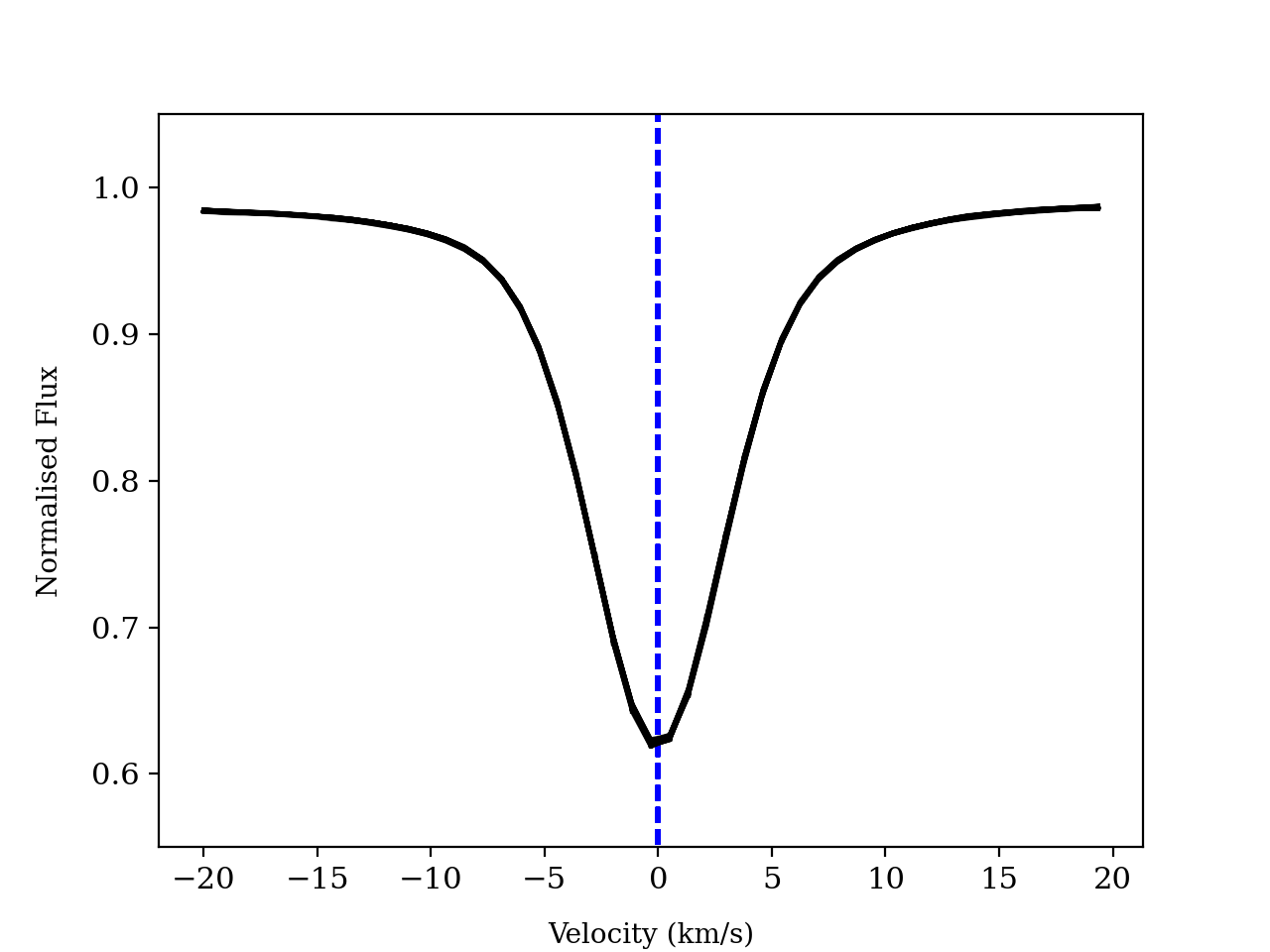}
    \caption{The CCF profiles (left) and final $\codename$ line profiles (right) for HD189733b. The CCFs provided by HARPS were over-sampled by a factor of 4, for easier comparison, every 4\textsuperscript{th} element in the CCFs has been plotted. All frames over all 4 observation nights are plotted on top of each other. This shows uniformity in both CCF and $\codename$ profiles between the observation nights. A vertical dashed blue line has been drawn through 0\,km\,s$^{-1}$ to help highlight the shape of the profiles.}
    \label{fig:full_prof}
\end{figure*}



\subsection{\codename ~Profiles}
\label{sec:profiles}

\indent The application of $\codename$ was done separately for each observation night to account for any night-to-night variations in the spectral continuum. The final $\codename$ profiles and their comparative CCFs are presented for all frames and all observation nights in Fig.~\ref{fig:full_prof}. The CCFs provided by HARPS are over-sampled by a factor of 4. This oversampling was present in DRS pipeline version 3.5 and was removed from subsequent pipelines due to associated issues such as red noise. We choose not to over-sample our $\codename$ profiles to avoid masking the true S/N of the profiles and hence for easier comparison, every 4\textsuperscript{th} element in the CCFs has been plotted in Fig.~\ref{fig:full_prof}. We also adjusted the continua of the $\codename$ profiles to match that of the CCFs by adding 1 to the profiles.
\newline \indent Error bars are included for the $\codename$ profiles but are barely visible due to the small errors associated with the profiles. These errors are representative of the data as LSD carries out a direct fit to the data. Error bars are not included for CCFs as they are not provided by the HARPS pipeline used for these observations. As well as producing realistic and representative errors, $\codename$ profiles are also seen to rival the signal-to-noise levels of the CCFs. 
\newline \indent It may be noticed that the continuum of the $\codename$ profiles generally sits lower than that of the CCFs. The explanation for this is two-fold. First, the continuum of the CCFs has been artificially improved by normalising the CCFs after their extraction from the spectrum. This normalisation is done by dividing the CCF by the mean continuum flux, hence the continuum would be brought to 1. Second, as mentioned in Section \ref{sec:LSD_tech}, very small absorption lines sitting below the noise level of the spectrum were excluded from the line list to save computational time. Despite these lines being undetectable, i.e. their profiles are entirely enveloped in the noise of the continuum, the collective effect of these lines can still cause the continuum of the profile to be suppressed. It is important to note that these lines (and their combined effect) lie in the noise of the spectral continuum and hence including them would not affect the continuum fit found by $\codename$. Consequently, although a similar offset is observed in the continuum of the profile this effect does not induce the same line distortions and increased noise levels associated with an incorrect continuum correction outlined in Section \ref{sec:Stellar Continuum Correction}. 
\begin{figure*}
        \includegraphics[scale = 0.6, trim = {0, 0, 1.1cm, 1.4cm}, clip]{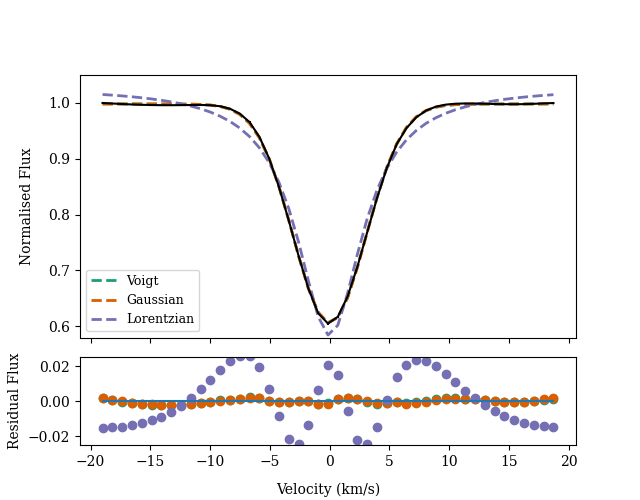}
	\includegraphics[scale = 0.6, trim = {2cm, 0, 1cm, 1.4cm}, clip]{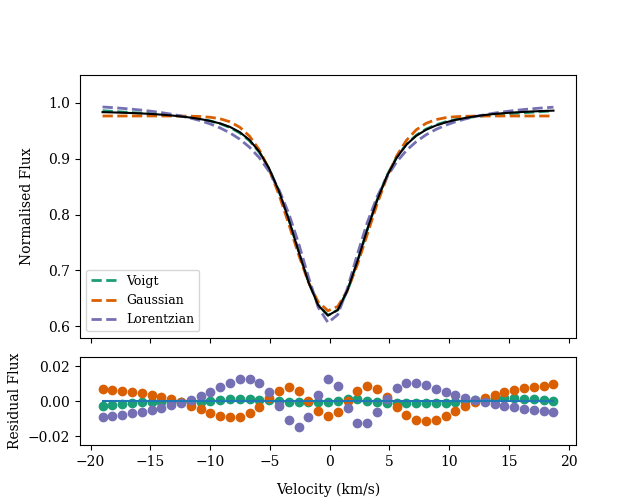}
    \caption{The top panels show the average out-of-transit CCF (left) and $\codename$ profile (right) for all nights are plotted as a solid black line. The best fit Voigt, Gaussian and Lorentzian profiles are plotted. It can be seen that the CCF exhibits a very Gaussian shape, while the $\codename$ profile is best described by a Voigt profile. The residuals are shown in the bottom panel.}
    \label{fig:profile_shape}
\end{figure*}

\indent Furthermore, the CCFs and $\codename$ profiles also exhibit distinctly different shapes, which we now discuss. Of particular note is the fact that the $\codename$ profiles have depths that are considerably more consistent than those of the CCFs across all 4 observation nights. Indeed, the standard deviation of the normalised $\codename$ profile depths was measured as 0.0003, while for the CCFs this was 0.0012 -- four times greater. Variations in the profile depths can be caused by an incorrect continuum fit as this results in systematic errors in the resulting LSD profile continuum as well. The uniformity between the $\codename$ profiles' continua and depths across all frames is supportive that $\codename$ has correctly identified and removed the spectral continuum before LSD is applied. 
\newline \indent The $\codename$ profiles also exhibit broader wings that tail off gradually to form the profile's continuum. This is contrasted by the well-defined transition between the profile and continuum that is seen for the CCFs. The shape of a stellar line profile is determined by the broadening effects that affect the light coming from the star. The absorption coefficient, a measure of how the stellar atmosphere absorbs light, is dominated by 3 main broadening processes: thermal, natural and pressure broadening. Thermal broadening arises from the movement of atoms due to their increased energy with higher temperatures and is expected to produce a Gaussian-shaped profile. Natural and pressure broadening are expected to produce a Lorentzian-shaped profile, possessing the characteristic broad wings as seen in the $\codename$ profiles. In general, the absorption coefficient should therefore account for a combination of these broadening processes, this would result in a dispersion profile in the shape of the Voigt profile, which describes the convolution between a Lorentzian and Gaussian profile, possessing a Gaussian line core with Lorentzian wings. Although the absorption coefficient will influence the shape of the stellar line profile, other broadening effects such as rotational, macroturbulence and instrumental broadening also affect the observed line profile shape. Instrumental broadening occurs due to limitations in the resolution of the instrument and is usually approximated by a Gaussian. Rotational broadening occurs from the rotation of the star as the velocity along the line of sight varies for different surface regions. Macroturbulence broadening refers to the broadening that occurs from large-scale mass motions and the Doppler shifts they induce \citep{Gray_stellar}.
\newline \indent In the case of optically thick lines, the line profile is expected to become broader and deeper as it originates from denser regions of the stellar atmosphere \citep{Gray_stellar} and since $\codename$ works in effective optical depth this effect should be accounted for in the $\codename$ profiles. HARPS CCFs are constructed using a weighted sum of the spectral lines using a binary mask. This means that, in the same way as the $\codename$ profiles, the CCFs should be representative of all of these broadening effects previously mentioned.
\newline \indent To investigate the shape of the HARPS CCFs compared to the $\codename$ profiles model Lorentzian, Gaussian and Voigt profiles were fit to the average out-of-transit profile in each case. The best-fit model for the CCF and $\codename$ profile can be seen in Fig.~\ref{fig:profile_shape} as well as the average out-of-transit profile and CCF taken across all 4 observation nights. It can be seen that the $\codename$ profiles are best fit by a Voigt model and the CCFs are best fit by a Gaussian model. The Voigt profile provides a better approximation to the expected shape of the line profile as it describes the shape of the line absorption coefficient. While the line absorption coefficient cannot be used as a direct comparison to the final shape of the line profile it would contribute to this. These combined suggest that $\codename$ is producing more realistic line profiles than the CCFs. A possible explanation for this comes from how the CCFs are constructed. A HARPS CCF is constructed by cross-correlating the spectrum with a stellar mask. Lines that are not removed by the mask are then added together to form the final CCF. This process does not account for the blending of neighbouring lines and in the case of many shallow lines, these can blend together to form a pseudo continuum, sitting lower than the true continuum. This would also cause the wings of the deeper lines to be lost in this pseudo-continuum. This could explain the Gaussian shape of the CCF as it only represents the core of the line, rather than the entire profile. As $\codename$ utilises LSD it accounts for the overlapping of neighbouring lines and is not subject to this effect, hence producing the Voigt shape seen for the $\codename$ profiles.

\begin{figure}	
    \includegraphics[width = \columnwidth, trim = {0, 0, 0, .8cm}, clip]{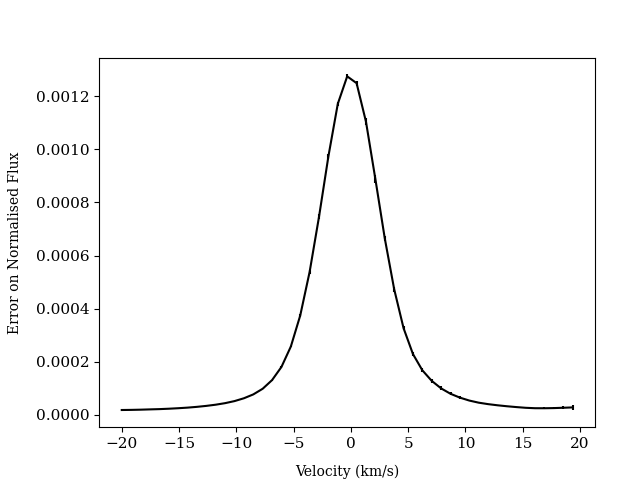}
    \caption{Average errors from $\codename$ profiles across all nights. Error bars shown represent the standard deviation in the errors.}
    \label{fig:profile_errors}
\end{figure}

\begin{figure}	
    \includegraphics[width = \columnwidth, trim = {0cm, 1.6cm, 1.5cm, 4cm}, clip]{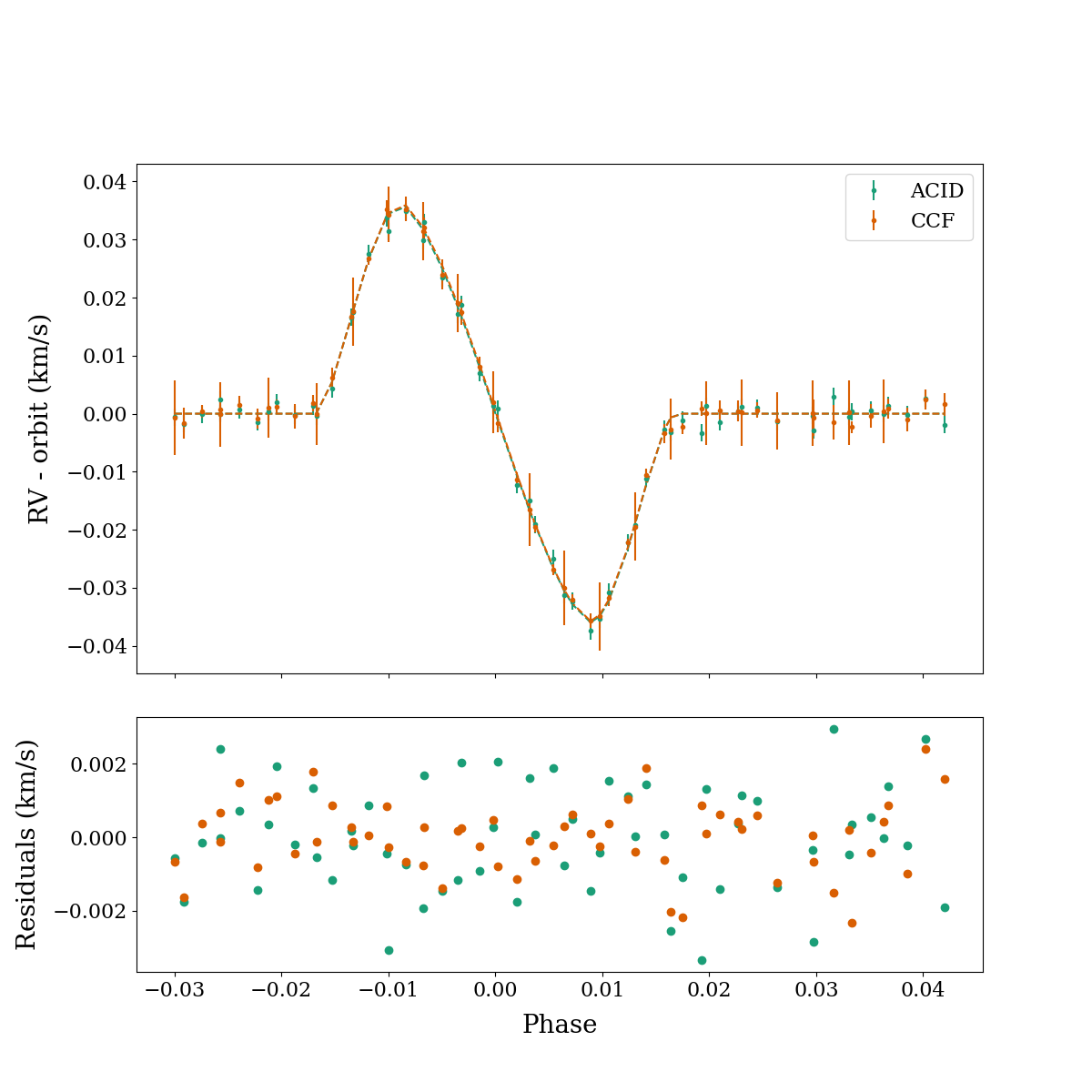}
    \caption{The top panel shows the radial velocities (RVs) as a function of phase for the CCFs and $\codename$ profiles for HD189733b. The RVs were taken as the mean from a best-fit Gaussian and Voigt model for the CCFs and $\codename$ profiles, respectively. Best-fit RM models are shown by dashed lines in each case. The residuals are shown in the bottom panel.}
    \label{fig:full_rvs}
\end{figure}


\begin{figure*}
	\includegraphics[scale = 0.26, trim={0cm, 0, 5.9cm, 0}, clip]{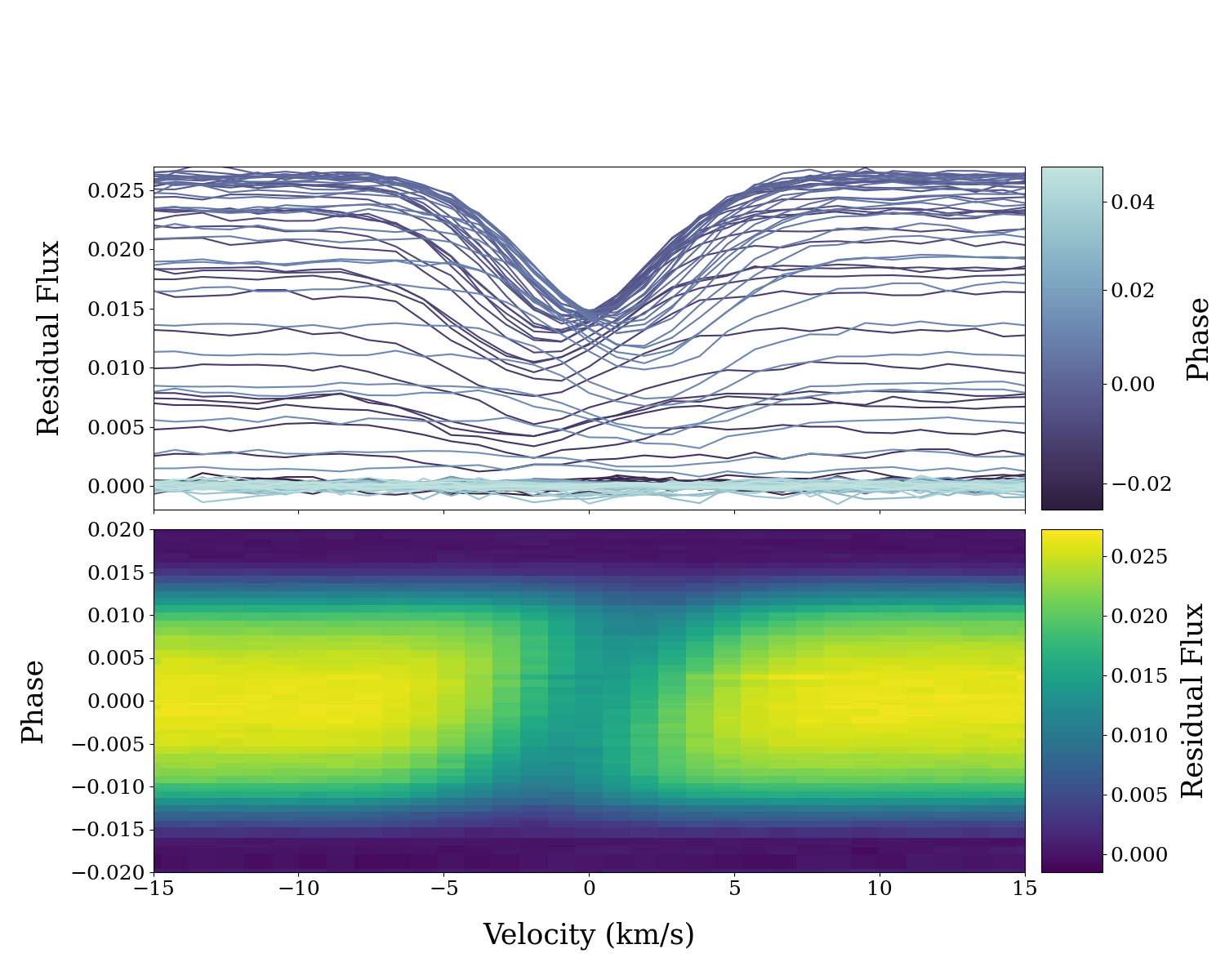} 
        \includegraphics[scale = 0.26, trim={3.3cm, 0, 0cm, 0}, clip]{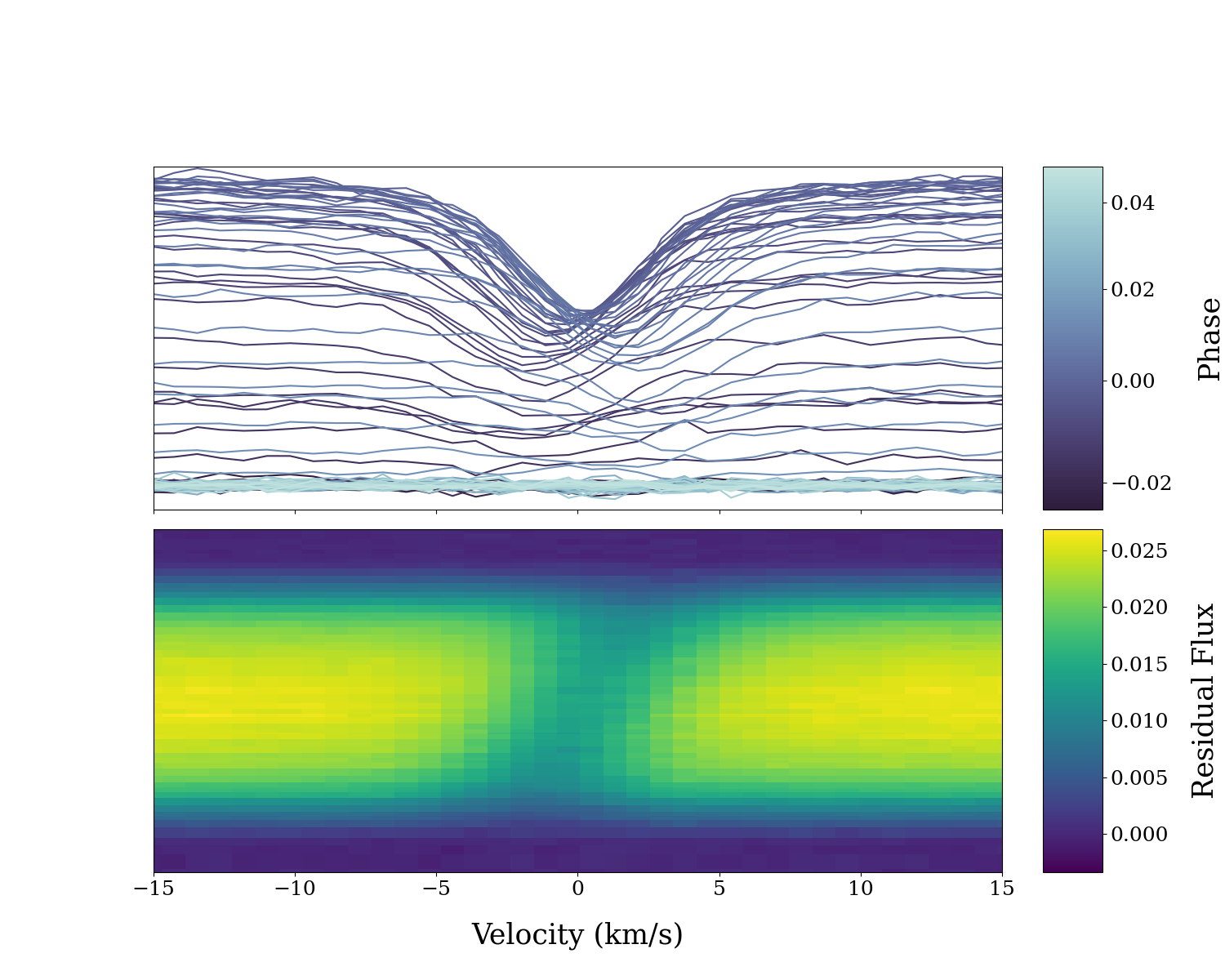}
    \caption{The residual profiles, obtained via the Reloaded RM (RRM) method from HARPS CCFs (left - all processed using the same K5 mask) and $\codename$ profiles (right) are shown in the top panel for each case. Their corresponding phase maps are shown in the bottom panel. The isolated RM signal can be seen as a travelling `bump' in the residual profiles and as a dark streak travelling across the centre of the phase map. The CCFs are plotted using every 4th element to account for the oversampling from the HARPS pipeline.}
    \label{fig:ccfs}
\end{figure*}

\begin{figure}	
    \includegraphics[scale = 0.55, trim = {0cm, 0cm, 1cm, 1cm}, clip]{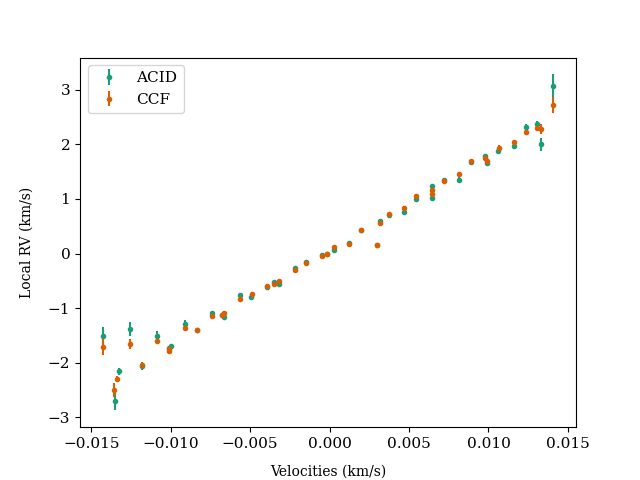}
    \caption{Net velocity shifts of the in-transit residual profiles as a function of phase for the CCF residual profiles and $\codename$ residual profiles. }
    \label{fig:resi_rvs}
\end{figure}

Other stellar effects can also affect the profile shape. For example, granulation on the stellar surface, where rising (i.e. blue-shifted) regions of the star are hotter and so brighter than the down-flowing (red-shifted) cooler regions, produces an asymmetry in the stellar line profile. In this case, the blue side of the profile is brighter and the red side of the profile possesses a stronger `tail' off to the continuum. The extent of this asymmetry is expected to vary between different stellar lines as they originate from different depth ranges in the stellar atmosphere and therefore sample different regions of the convective motions. We attempted to measure the asymmetry of the $\codename$ profiles by fitting a skew-normal model (e.g~\citealt{2019A&A...622A.131S}) that fits the line using a `skewed' Gaussian profile and hence can quantify a line-profile's asymmetry. As can be seen in Fig.~\ref{fig:profile_shape}, the $\codename$ profiles are not well described by a Gaussian model. Therefore, the addition of the skewness parameter in the `skewed' Gaussian model prevented the model fit from converging and hence this test yielded inconclusive results. Asymmetry was further investigated by calculating the bisector for the average out-of-transit CCF and $\codename$ profile. The bisector was found for a normalised flux range of 0.65 - 0.95 at 0.01 intervals. The bisector span for both the CCFs ($0.11\pm0.07$~km s$^{-1}$) and $\codename$ profiles ($0.07\pm0.07$~km s$^{-1}$) do not suggest an obvious asymmetry in either the CCFs or $\codename$ profiles. 
\newline \indent It should be noted, however, that the $\codename$ profiles do exhibit some asymmetry in the errors produced. Fig.~\ref{fig:profile_errors} shows the profile errors for all frames as a function of velocity. It can be seen that the errors increase in the centre of the profile as expected (due to the lower signal-to-noise in the line cores), but they also appear to be larger on the red wing of the profile relative to the blue wing. This asymmetry was not observed when $\codename$ was applied to model data (which assumed symmetric line-profile shapes) and so is likely originating from the observed spectra. One possible explanation for this is that it arises from granulation, which will cause different lines to exhibit different levels of line-asymmetries (due to, for example, differences in line-depth formation), with the greatest line-by-line variation occurring in the red wing of the profile. This would lead to shape variations between lines that were predominantly larger in the red wings, hence leading to the larger uncertainties seen in this region of the $\codename$ `mean' line profiles.

\subsection{RM Analysis}
\label{sec: RM_analysis}

\indent The resulting $\codename$ profiles were used to measure the classical RM effect of HD189733b by measuring the RVs of the profiles. RVs were taken to be the mean of a Voigt and Gaussian fit to the $\codename$ profiles and CCFs, respectively; these are shown in Fig.~\ref{fig:full_rvs}. The best-fit model was found using a least$-$squares Levenberg-Marquardt algorithm implemented by \texttt{scipy's} \texttt{curvefit} function. The errors on the mean are taken from the square root of the diagonal of the covariance matrix returned. The CCF errors were taken as the standard deviation in the CCF's continuum, while for the $\codename$ profiles the corresponding profile errors were used. In both cases, areas of the continuum that deviated from the model were excluded. For the CCFs this meant points lying outside of an absolute velocity of 5\,km\,s$^{-1}$ were excluded and for $\codename$ profile points lying outside of an absolute velocity of 7\,km\,s$^{-1}$ were excluded. A circular orbit was assumed for HD189733b and hence the RV motion of the host star was found using the best-fit model of a phase-dependent sinusoid with semi-amplitude, $K$, plus a systemic radial velocity, $v_{\mathrm{sys}}$. This reflex motion was removed from the measured RVs. In general, the RVs measured from the $\codename$ profiles (see Fig.~\ref{fig:full_rvs}) show a slightly increased scatter compared to those from the CCFs with an RMS of the out-of-transit RVs being 0.0015\,km\,s$^{-1}$ and 0.0011\,km\,s$^{-1}$  for the $\codename$ profiles and CCFs, respectively. RV scatter can be induced in LSD profiles from a number of sources including unmasked telluric lines, overlapping neighbouring lines and insufficiencies in the model (e.g. missing or inaccurate lines in the line list). Most of these cases will have been accounted for in the masking stage of $\codename$, when known telluric regions and wavelength regions that greatly deviated from the model were removed (Section \ref{sec:masking}). It was seen in Section \ref{sec:LSD_od} that although $\codename$ produces a better fit to the data through applying LSD in effective optical depth discrepancies remained between the data and model (Fig.~ \ref{fig:flux_od_LSD}). This discrepancy is a possible source of the increased RMS seen in the RVs from $\codename$ profiles compared to those from the CCFs. 

\begin{table}
	\centering
	\begin{tabular}{ccc}
            \hline
            \textbf{Parameter} & \textbf{Value} & \textbf{Reference}\\
		\hline
            $\sigma_{a}$ & $3.22\pm0.04$~km s$^{-1}$ & - \\
            $d_{a}$ & $0.369\pm0.004$ & - \\
            $\sigma_{c}$ & $3.038\pm0.004$~km s$^{-1}$ & - \\
            $d_{c}$ & $0.3927\pm0.0003$ & - \\
            $v\sin i_{*}$ & $3.316\pm^{0.017}_{0.067}$~km s$^{-1}$ & {\cite{2009A&A...506..377T}} \\
            $i_{p}$ & $85.465\pm0.002\degree$ & {\cite{2023arXiv231006681C}} \\
            $R_{p}/R_{*}$ & $0.15667\pm{0.00012}$ & {\cite{2009A&A...506..377T}} \\
            $a/R_{*}$ & $8.7866\pm{0.0082}$ & {\cite{2023arXiv231006681C}} \\
            ($u_{1}$, $u_{2}$) & ($0.548$, $0.213$) & {\cite{2016A&A...588A.127C}} \\
            $P$ & $2.21857567$ days & {\cite{2010ApJ...721.1861A}} \\
		\hline
	\end{tabular}
        \vspace{.5cm}
    \caption{Planetary and stellar parameters used for modelling the RM RV curve. The inputs for the model are: Gaussian half-width for $\codename$, $\sigma_{a}$ and CCFs, $\sigma_{c}$ ; normalised profile depth for $\codename$, $d_{a}$ and CCFs, $d_{c}$; stellar rotational velocity, $v\sin i_{*}$; planetary inclination, $i_{p}$; effective planetary radius, $R_{p}/R_{*}$; scaled semi-major axis, $a/R_{*}$; limb darkening coefficients, ($u_{1}$, $u_{2}$) and the orbital period, $P$.}
    \label{tab:RM_inputs}
\end{table}

\indent The RM curve was modelled using the method outlined in \cite{2013A&A...550A..53B} with quadratic limb darkening applied. Planetary and stellar inputs, except the projected spin-orbit angle, $\lambda$, were fixed to values obtained from the literature (outlined in Table \ref{tab:RM_inputs}). Uniform priors were applied to the profile depth and width based on the best-fit models and uncertainties outlined in Section \ref{sec:profiles}. These corresponded to the Gaussian half-width and depth. The width of each of the priors was equal to the uncertainty on the best-fit value in each case. Observations taken in July 2006 covered a partial transit only as bad weather conditions were reported. Furthermore, \cite{2016A&A...588A.127C} reported deviations in the local RVs measured from the residual profiles from July 2007. It was suggested that this deviation was caused by a possible decrease in stellar activity on this night compared to the other observation nights. We also saw this deviation in our residual profiles (discussed later in this section) and hence chose to exclude this observation night from the RM fit. Following this, the model was only applied to data from August 2007 and September 2006 and yields a $\lambda$ = $0.05 \pm 0.43 \degree$ and $\lambda$ = $0.13\pm0.34\degree$ for the CCFs and $\codename$ profiles, respectively. In both cases, the spin-orbit angle agrees within the uncertainties with previously reported values ($-1.7\pm0.8\degree$ to $-0.35\pm0.25\degree$) \citep{2016A&A...588A.127C, 2009A&A...506..377T, 2010MNRAS.403..151C}. It can be noted that $\codename$ produces a similar constraint on the spin-orbit angle despite the increased scatter in the RV measurements. This demonstrates $\codename$'s ability to detect planetary effects in the RVs to the same precision as CCFs. 
\newline \indent The RM effect of HD189733b was further investigated using the Reloaded RM (RRM) technique. This technique, introduced by \cite{2016A&A...588A.127C}, isolates the local stellar line profile occulted by the planet. An in-depth description of this technique is included in \cite{2016A&A...588A.127C}, however a brief summary will follow. The line profiles are first scaled according to a Mandel and Agol transit light curve with a quadratic limb darkening law (using the limb darkening coefficients outlined in Table \ref{tab:RM_inputs}). The RRM technique then relies on creating a master out-of-transit profile by combining all out-of-transit profiles using a weighted mean. A residual profile for a particular frame is then created by subtracting the line profile of that frame from the master out-of-transit profile. In this way, the RM signal is isolated and can be seen as a travelling `bump' in the phase map of the residual profiles. This analysis applies the RRM technique to observations from September 2006, July 2007 and August 2007 as was done by \cite{2016A&A...588A.127C}. This is done on a night-by-night basis for both $\codename$ profiles and CCFs to account for any small variations that could arise in the master out-of-transit profile between each night. 
\newline \indent The CCFs used in our analysis are those produced from the same HARPS pipeline as \cite{2016A&A...588A.127C}. However, the HARPS CCFs available on the archive for these observation nights are processed using different stellar masks for different nights. Namely, a K5 stellar mask was applied for observations taken in September 2006 and a G2 stellar mask was applied for July and August 2007. We observed variations in the depths of the CCFs dependent on which mask was used. As this made it difficult to compare the CCFs and $\codename$ profiles we requested the CCFs for all observation nights to be reprocessed using only the K5 stellar mask which provided a better match to the spectral type of HD189733 (K2). This ensured the depths of the CCFs were relatively constant across the observation nights. The residual profiles and CCFs for HD189733b are shown in the top panel of Fig.~\ref{fig:ccfs} with their corresponding phase maps shown in the bottom panel. The RRM signal can be seen as a travelling `bump' visible in residual profiles and CCFs and a dark streak in the phase maps. 
The signal-to-noise of the CCFs compared to the $\codename$ profiles is very similar, with the RMS of the out-of-transit residual profiles being 0.0012 and 0.0013 in normalised flux for $\codename$ profiles and CCFs, respectively.  
\newline \indent The RVs of the residual profiles, representing the local RV of the stellar surface blocked by the planet, were also measured using the mean of a Gaussian (for the CCFs) or Voigt (for the $\codename$ profiles) fit to the in-transit residual profiles. To avoid the continuum of the residual $\codename$ profiles and CCFs, which generally deviate from the models, points lying outside of an absolute velocity of 10 km s$^{-1}$ were excluded. A larger velocity range was included for the residual profile fit compared to that for the full profiles to account for the larger RV shifts. As the HARPS CCF profiles do not contain errors the error was taken as the standard deviation in the CCF's continuum while for the $\codename$ profiles, the corresponding profile errors were propagated through for the residual profiles. Fig.~\ref{fig:resi_rvs} shows the residual RVs for both the CCFs and $\codename$ profiles. The RVs from $\codename$ have a slightly larger scatter and error associated with them. Despite this, it can be seen that for both the CCFs and $\codename$ profiles the local RV changes as the planet moves across the stellar disc. The first half of the planet's transit shows a blue-shifted stellar surface (moving towards the observer) with this switching to a red-shifted stellar surface (moving away from the observer) once the planet has crossed the centre of the stellar disc. This is the expected outcome as HD189733b resides in an aligned, prograde orbit around its star. Both CCFs and $\codename$ profiles exhibit velocities of $\sim$3\,km\,s$^{-1}$ at the stellar limbs. This agrees with the results presented in \cite{2016A&A...588A.127C} as well as the $v \sin{i_*}$ values reported in the literature, which range from $2.9-3.5$\,km\,s$^{-1}$ (see \citealt{2016A&A...588A.127C} and references therein). 
\newline \indent The limbs also exhibit an increase in scatter in the RVs of both the $\codename$ profiles and CCFs. This deviation in the RVs from the residual CCFs was also seen in \cite{2016A&A...588A.127C} who attributed this to a possible decrease in magnetic stellar activity during this night. 
$\codename$ was seen to match the local RVs obtained from the CCFs within the uncertainties and produced a similar RV scatter to that seen with the CCFs. $\codename$ can therefore rival the residual profile RV precision obtained by CCFs while having the added benefit over CCFs as it produces profiles that better represent the expected stellar line profile shape (see Section \ref{sec:profiles}).

\section{Conclusions}
\label{sec:conclusion} 

This paper has successfully shown $\codename$ to be a novel technique that produces high precision LSD line profiles from high-resolution spectra. $\codename$ was seen to remedy LSD's prediction of nonphysical line depths and produce a forward model that better agreed with the data by performing LSD in effective optical depth. We saw a 40-50 times increase in the out-of-line RMS and an increased distortion of the line profile shape of the resulting LSD profile when an incorrect continuum fit is applied, highlighting the importance of an accurate continuum fit. $\codename$ applies a combined fit of the mean line profile and the spectral continuum which, when applied to model data, was able to accurately retrieve the `true' spectral continuum. We have shown that implementing this approach provides an accurate continuum correction and, in turn, minimises the noise and profile distortions that can be induced into the resulting profiles. 
\newline \indent The application of $\codename$ on HARPS data for HD189733b produced high S/N LSD profiles that exhibit a Voigt profile shape. This is a better approximation for the `true' profile shape as it shows evidence for a combination of broadening effects. In the case of the CCFs, we observed a Gaussian shape, therefore not accounting for all broadening effects that can occur. $\codename$, therefore, better preserved the shape of the `true' line profile and hence the encoded physics than the corresponding CCFs. While neither the $\codename$ profiles nor the CCFs showed evidence of asymmetry in the shape of line profiles, an asymmetry was observed in the uncertainties of the $\codename$ profiles. This could be indicative of line-by-line variability in the asymmetrical shape induced by stellar granulation.
\newline \indent Furthermore, $\codename$ was shown to accurately produce the RM signal of HD189733b through both the classical RM technique and Reloaded RM. For the classical RM technique, we saw that $\codename$ was able to match the RVs obtained from HARPS CCFs within the uncertainties. We did however observe a slightly increased scatter compared to that of the CCFs with an out-of-transit RV RMS of 0.0011 and 0.0015\,km\,s$^{-1}$ for the CCFs and $\codename$ profiles, respectively. RV scatter in the LSD profiles can be induced by the discrepancies between the data and the model. We observed that although $\codename$ reduces these discrepancies by performing LSD in effective optical depth, some remained as a result of inaccuracies in the line list and possibly remaining `incorrect' assumptions in LSD. We would therefore expect the RV scatter to decrease when more accurate line data can be obtained. Additionally, the RV scatter could be further improved by also improving the model used to extract the RVs. For example, accounting for any asymmetry in the profile shape that could arise from stellar granulation on the star's surface would improve the accuracy of the measured RVs. While we did not observe evidence of asymmetry in the $\codename$ profiles for HD189733, there was potential evidence of asymmetry found in the profile errors. \cite{2019A&A...622A.131S} has introduced a `skewed' Gaussian profile that models this asymmetry. We saw that $\codename$ profiles exhibit a Voigt profile shape and were not modelled well by a Gaussian profile. Hence a combination of this `skewed' model with a Voigt profile shape would be required to better measure the RVs of the $\codename$ profiles.
\newline \indent Best fit RM models applied to the CCFs and $\codename$ profiles returned a spin-orbit angle of $\lambda$ = $0.05 \pm 0.43 \degree$ and $\lambda$ = $0.13 \pm 0.34\degree$ for the CCFs and $\codename$ profiles, respectively. This showed that a 0.4\,m\,s$^{-1}$ increase in the out-of-transit RV RMS is small enough that it did not affect the precision to which $\codename$ could retrieve the spin-orbit angle. The application of the Reloaded RM technique to both $\codename$ profiles and HARPS CCFs shows $\codename$ residual profiles improve the out-of-line RMS by over $5\%$ compared to CCFs while maintaining the RV precision in the local RVs measured from the residual profiles.
This further demonstrates that $\codename$ can match the precision of the CCFs while providing a better representation of the line profile. This capability of delivering high S/N profiles that accurately represent the stellar line profile will contribute to our understanding of stellar and planetary physics by providing better insights into their effect on the stellar line profile.


\section*{Acknowledgements}

This research has made use of NASA’s Astrophysics Data
System Bibliographic Services, and the VALD database, operated at Uppsala
University, the Institute of Astronomy RAS in Moscow, and the University of Vienna.  C.A.W. and E.dM. would like to acknowledge support from the UK Science and Technology Facilities Council (STFC, grant number ST/X00094X/1).
L.S.D would like to acknowledge funding and support from the UK Science and Technology Facilities Council (STFC grant number ST/V505990/1).
We would also like to thank the anonymous reviewer for their useful comments.

\section*{Data Availability}

Based on observations collected at the European Organisation for Astronomical Research in the Southern Hemisphere under ESO programmes 072.C-04889(E), 079.C-0828(A) and 079.C-0127(A), and processed data created thereof. 
\newline The $\codename$ python code can be accessed and downloaded from the GitHub repository: \href{https://github.com/ldolan05/ACID.}{https://github.com/ldolan05/ACID.}
\newline Documentation and installation instructions available at: \href{https://acid-code.readthedocs.io/en/latest/index.html}{https://acid-code.readthedocs.io/en/latest/index.html}




\bibliography{biblio.bib} 

\begin{thebibliography}{}
\makeatletter
\relax
\def\mn@urlcharsother{\let\do\@makeother \do\$\do\&\do\#\do\^\do\_\do\%\do\~}
\def\mn@doi{\begingroup\mn@urlcharsother \@ifnextchar [ {\mn@doi@}
  {\mn@doi@[]}}
\def\mn@doi@[#1]#2{\def\@tempa{#1}\ifx\@tempa\@empty \href
  {http://dx.doi.org/#2} {doi:#2}\else \href {http://dx.doi.org/#2} {#1}\fi
  \endgroup}
\def\mn@eprint#1#2{\mn@eprint@#1:#2::\@nil}
\def\mn@eprint@arXiv#1{\href {http://arxiv.org/abs/#1} {{\tt arXiv:#1}}}
\def\mn@eprint@dblp#1{\href {http://dblp.uni-trier.de/rec/bibtex/#1.xml}
  {dblp:#1}}
\def\mn@eprint@#1:#2:#3:#4\@nil{\def\@tempa {#1}\def\@tempb {#2}\def\@tempc
  {#3}\ifx \@tempc \@empty \let \@tempc \@tempb \let \@tempb \@tempa \fi \ifx
  \@tempb \@empty \def\@tempb {arXiv}\fi \@ifundefined
  {mn@eprint@\@tempb}{\@tempb:\@tempc}{\expandafter \expandafter \csname
  mn@eprint@\@tempb\endcsname \expandafter{\@tempc}}}

\bibitem[\protect\citeauthoryear{{Agol}, {Cowan}, {Knutson}, {Deming},
  {Steffen}, {Henry}  \& {Charbonneau}}{{Agol}
  et~al.}{2010}]{2010ApJ...721.1861A}
{Agol} E.,  {Cowan} N.~B.,  {Knutson} H.~A.,  {Deming} D.,  {Steffen} J.~H.,
  {Henry} G.~W.,   {Charbonneau} D.,  2010, \mn@doi [\apj]
  {10.1088/0004-637X/721/2/1861}, \href
  {https://ui.adsabs.harvard.edu/abs/2010ApJ...721.1861A} {721, 1861}

\bibitem[\protect\citeauthoryear{{Barnes} et~al.,}{{Barnes}
  et~al.}{2012}]{2012MNRAS.424..591B}
{Barnes} J.~R.,  et~al., 2012, \mn@doi [\mnras]
  {10.1111/j.1365-2966.2012.21236.x}, \href
  {https://ui.adsabs.harvard.edu/abs/2012MNRAS.424..591B} {424, 591}

\bibitem[\protect\citeauthoryear{{Bou{\'e}}, {Montalto}, {Boisse}, {Oshagh}  \&
  {Santos}}{{Bou{\'e}} et~al.}{2013}]{2013A&A...550A..53B}
{Bou{\'e}} G.,  {Montalto} M.,  {Boisse} I.,  {Oshagh} M.,   {Santos} N.~C.,
  2013, \mn@doi [\aap] {10.1051/0004-6361/201220146}, \href
  {https://ui.adsabs.harvard.edu/abs/2013A&A...550A..53B} {550, A53}

\bibitem[\protect\citeauthoryear{{Boyajian} et~al.,}{{Boyajian}
  et~al.}{2015}]{2015MNRAS.447..846B}
{Boyajian} T.,  et~al., 2015, \mn@doi [\mnras] {10.1093/mnras/stu2502}, \href
  {https://ui.adsabs.harvard.edu/abs/2015MNRAS.447..846B} {447, 846}

\bibitem[\protect\citeauthoryear{{Casagrande}, {Sch{\"o}nrich}, {Asplund},
  {Cassisi}, {Ram{\'\i}rez}, {Mel{\'e}ndez}, {Bensby}  \&
  {Feltzing}}{{Casagrande} et~al.}{2011}]{2011A&A...530A.138C}
{Casagrande} L.,  {Sch{\"o}nrich} R.,  {Asplund} M.,  {Cassisi} S.,
  {Ram{\'\i}rez} I.,  {Mel{\'e}ndez} J.,  {Bensby} T.,   {Feltzing} S.,  2011,
  \mn@doi [\aap] {10.1051/0004-6361/201016276}, \href
  {https://ui.adsabs.harvard.edu/abs/2011A&A...530A.138C} {530, A138}

\bibitem[\protect\citeauthoryear{{Cegla}, {Lovis}, {Bourrier}, {Beeck},
  {Watson}  \& {Pepe}}{{Cegla} et~al.}{2016}]{2016A&A...588A.127C}
{Cegla} H.~M.,  {Lovis} C.,  {Bourrier} V.,  {Beeck} B.,  {Watson} C.~A.,
  {Pepe} F.,  2016, \mn@doi [\aap] {10.1051/0004-6361/201527794}, \href
  {https://ui.adsabs.harvard.edu/abs/2016A&A...588A.127C} {588, A127}

\bibitem[\protect\citeauthoryear{Collier~Cameron}{Collier~Cameron}{2001}]{collier2001}
Collier~Cameron A.,  2001, in Boffin H. M.~J.,  Steeghs D.,  eds, Lecture Notes
  in Physics, Astrotomography: Indirect Imaging Methods in Observational
  Astronomy.
Springer-Verlag, Berlin, p.~183

\bibitem[\protect\citeauthoryear{{Collier Cameron}, {Bruce}, {Miller}, {Triaud}
   \& {Queloz}}{{Collier Cameron} et~al.}{2010}]{2010MNRAS.403..151C}
{Collier Cameron} A.,  {Bruce} V.~A.,  {Miller} G.~R.~M.,  {Triaud}
  A.~H.~M.~J.,   {Queloz} D.,  2010, \mn@doi [\mnras]
  {10.1111/j.1365-2966.2009.16131.x}, \href
  {https://ui.adsabs.harvard.edu/abs/2010MNRAS.403..151C} {403, 151}

\bibitem[\protect\citeauthoryear{{Cristo} et~al.,}{{Cristo}
  et~al.}{2023}]{2023arXiv231006681C}
{Cristo} E.,  et~al., 2023, \mn@doi [arXiv e-prints]
  {10.48550/arXiv.2310.06681}, \href
  {https://ui.adsabs.harvard.edu/abs/2023arXiv231006681C} {p. arXiv:2310.06681}

\bibitem[\protect\citeauthoryear{{Donati}, {Semel}, {Carter}, {Rees}  \&
  {Collier Cameron}}{{Donati} et~al.}{1997}]{1997MNRAS.291..658D}
{Donati} J.~F.,  {Semel} M.,  {Carter} B.~D.,  {Rees} D.~E.,   {Collier
  Cameron} A.,  1997, \mn@doi [\mnras] {10.1093/mnras/291.4.658}, \href
  {https://ui.adsabs.harvard.edu/abs/1997MNRAS.291..658D} {291, 658}

\bibitem[\protect\citeauthoryear{{Esposito} et~al.,}{{Esposito}
  et~al.}{2014}]{2014A&A...564L..13E}
{Esposito} M.,  et~al., 2014, \mn@doi [\aap] {10.1051/0004-6361/201423735},
  \href {https://ui.adsabs.harvard.edu/abs/2014A&A...564L..13E} {564, L13}

\bibitem[\protect\citeauthoryear{{Fabrycky} \& {Tremaine}}{{Fabrycky} \&
  {Tremaine}}{2007}]{2007ApJ...669.1298F}
{Fabrycky} D.,  {Tremaine} S.,  2007, \mn@doi [\apj] {10.1086/521702}, \href
  {https://ui.adsabs.harvard.edu/abs/2007ApJ...669.1298F} {669, 1298}

\bibitem[\protect\citeauthoryear{{Foreman-Mackey}, {Hogg}, {Lang}  \&
  {Goodman}}{{Foreman-Mackey} et~al.}{2013}]{2013PASP..125..306F}
{Foreman-Mackey} D.,  {Hogg} D.~W.,  {Lang} D.,   {Goodman} J.,  2013, \mn@doi
  [\pasp] {10.1086/670067}, \href
  {https://ui.adsabs.harvard.edu/abs/2013PASP..125..306F} {125, 306}

\bibitem[\protect\citeauthoryear{{Goldreich} \& {Sari}}{{Goldreich} \&
  {Sari}}{2003}]{2003ApJ...585.1024G}
{Goldreich} P.,  {Sari} R.,  2003, \mn@doi [\apj] {10.1086/346202}, \href
  {https://ui.adsabs.harvard.edu/abs/2003ApJ...585.1024G} {585, 1024}

\bibitem[\protect\citeauthoryear{{Gray}}{{Gray}}{2022}]{Gray_stellar}
{Gray} D.~F.,  2022, {The Observation and Analysis of Stellar Photospheres.
  Fourth Edition}.
Cambridge University Press

\bibitem[\protect\citeauthoryear{{Grouffal} et~al.,}{{Grouffal}
  et~al.}{2022}]{2022A&A...668A.172G}
{Grouffal} S.,  et~al., 2022, \mn@doi [\aap] {10.1051/0004-6361/202244182},
  \href {https://ui.adsabs.harvard.edu/abs/2022A&A...668A.172G} {668, A172}

\bibitem[\protect\citeauthoryear{{Gustafsson}, {Edvardsson}, {Eriksson},
  {J{\o}rgensen}, {Nordlund}  \& {Plez}}{{Gustafsson} et~al.}{2008}]{atom_SME}
{Gustafsson} B.,  {Edvardsson} B.,  {Eriksson} K.,  {J{\o}rgensen} U.~G.,
  {Nordlund} {\r{A}}.,   {Plez} B.,  2008, \mn@doi [Astronomy and Astrophysics]
  {10.1051/0004-6361:200809724}, \href
  {https://ui.adsabs.harvard.edu/abs/2008A&A...486..951G} {486, 951}

\bibitem[\protect\citeauthoryear{{Jackson} et~al.,}{{Jackson}
  et~al.}{2023}]{2023MNRAS.518.4845J}
{Jackson} D.~G.,  et~al., 2023, \mn@doi [\mnras] {10.1093/mnras/stac3192},
  \href {https://ui.adsabs.harvard.edu/abs/2023MNRAS.518.4845J} {518, 4845}

\bibitem[\protect\citeauthoryear{{Kochukhov}, {Makaganiuk}  \&
  {Piskunov}}{{Kochukhov} et~al.}{2010}]{2010A&A...524A...5K}
{Kochukhov} O.,  {Makaganiuk} V.,   {Piskunov} N.,  2010, \mn@doi [\aap]
  {10.1051/0004-6361/201015429}, \href
  {https://ui.adsabs.harvard.edu/abs/2010A&A...524A...5K} {524, A5}

\bibitem[\protect\citeauthoryear{{Kozai}}{{Kozai}}{1962}]{1962AJ.....67..591K}
{Kozai} Y.,  1962, \mn@doi [\aj] {10.1086/108790}, \href
  {https://ui.adsabs.harvard.edu/abs/1962AJ.....67..591K} {67, 591}

\bibitem[\protect\citeauthoryear{{Kupka}, {Piskunov}, {Ryabchikova}, {Stempels}
   \& {Weiss}}{{Kupka} et~al.}{1999}]{1999A&AS..138..119K}
{Kupka} F.,  {Piskunov} N.,  {Ryabchikova} T.~A.,  {Stempels} H.~C.,   {Weiss}
  W.~W.,  1999, \mn@doi [\aaps] {10.1051/aas:1999267}, \href
  {https://ui.adsabs.harvard.edu/abs/1999A&AS..138..119K} {138, 119}

\bibitem[\protect\citeauthoryear{{Kupka}, {Ryabchikova}, {Piskunov}, {Stempels}
   \& {Weiss}}{{Kupka} et~al.}{2000}]{2000BaltA...9..590K}
{Kupka} F.~G.,  {Ryabchikova} T.~A.,  {Piskunov} N.~E.,  {Stempels} H.~C.,
  {Weiss} W.~W.,  2000, \mn@doi [Baltic Astronomy] {10.1515/astro-2000-0420},
  \href {https://ui.adsabs.harvard.edu/abs/2000BaltA...9..590K} {9, 590}

\bibitem[\protect\citeauthoryear{{Lidov}}{{Lidov}}{1962}]{1962P&SS....9..719L}
{Lidov} M.~L.,  1962, \mn@doi [\planss] {10.1016/0032-0633(62)90129-0}, \href
  {https://ui.adsabs.harvard.edu/abs/1962P&SS....9..719L} {9, 719}

\bibitem[\protect\citeauthoryear{{Lienhard}, {Mortier}, {Buchhave}, {Collier
  Cameron}, {L{\'o}pez-Morales}, {Sozzetti}, {Watson}  \&
  {Cosentino}}{{Lienhard} et~al.}{2022}]{2022MNRAS.513.5328L}
{Lienhard} F.,  {Mortier} A.,  {Buchhave} L.,  {Collier Cameron} A.,
  {L{\'o}pez-Morales} M.,  {Sozzetti} A.,  {Watson} C.~A.,   {Cosentino} R.,
  2022, \mn@doi [\mnras] {10.1093/mnras/stac1098}, \href
  {https://ui.adsabs.harvard.edu/abs/2022MNRAS.513.5328L} {513, 5328}

\bibitem[\protect\citeauthoryear{{Marsden} et~al.,}{{Marsden}
  et~al.}{2014}]{2014MNRAS.444.3517M}
{Marsden} S.~C.,  et~al., 2014, \mn@doi [\mnras] {10.1093/mnras/stu1663}, \href
  {https://ui.adsabs.harvard.edu/abs/2014MNRAS.444.3517M} {444, 3517}

\bibitem[\protect\citeauthoryear{{McLaughlin}}{{McLaughlin}}{1924}]{1924ApJ....60...22M}
{McLaughlin} D.~B.,  1924, \mn@doi [\apj] {10.1086/142826}, \href
  {https://ui.adsabs.harvard.edu/abs/1924ApJ....60...22M} {60, 22}

\bibitem[\protect\citeauthoryear{{Ol{\'a}h}, {Jurcsik}  \&
  {Strassmeier}}{{Ol{\'a}h} et~al.}{2003}]{2003A&A...410..685O}
{Ol{\'a}h} K.,  {Jurcsik} J.,   {Strassmeier} K.~G.,  2003, \mn@doi [\aap]
  {10.1051/0004-6361:20031352}, \href
  {https://ui.adsabs.harvard.edu/abs/2003A&A...410..685O} {410, 685}

\bibitem[\protect\citeauthoryear{{Pakhomov}, {Ryabchikova}  \&
  {Piskunov}}{{Pakhomov} et~al.}{2019}]{2019ARep...63.1010P}
{Pakhomov} Y.~V.,  {Ryabchikova} T.~A.,   {Piskunov} N.~E.,  2019, \mn@doi
  [Astronomy Reports] {10.1134/S1063772919120047}, \href
  {https://ui.adsabs.harvard.edu/abs/2019ARep...63.1010P} {63, 1010}

\bibitem[\protect\citeauthoryear{{Perryman}}{{Perryman}}{2018}]{exoplanet_handbook_2018}
{Perryman} M.,  2018, {The Exoplanet Handbook}.
Cambridge University Press

\bibitem[\protect\citeauthoryear{{Piskunov} \& {Valenti}}{{Piskunov} \&
  {Valenti}}{2017}]{2017SME}
{Piskunov} N.,  {Valenti} J.~A.,  2017, \mn@doi [\aap]
  {10.1051/0004-6361/201629124}, \href
  {https://ui.adsabs.harvard.edu/abs/2017A&A...597A..16P} {597, A16}

\bibitem[\protect\citeauthoryear{{Rainer} et~al.,}{{Rainer}
  et~al.}{2016}]{2016AJ....152..207R}
{Rainer} M.,  et~al., 2016, \mn@doi [\aj] {10.3847/0004-6256/152/6/207}, \href
  {https://ui.adsabs.harvard.edu/abs/2016AJ....152..207R} {152, 207}

\bibitem[\protect\citeauthoryear{{Ram{\'\i}rez V{\'e}lez}}{{Ram{\'\i}rez
  V{\'e}lez}}{2020}]{2020MNRAS.493.1130R}
{Ram{\'\i}rez V{\'e}lez} J.~C.,  2020, \mn@doi [\mnras]
  {10.1093/mnras/staa301}, \href
  {https://ui.adsabs.harvard.edu/abs/2020MNRAS.493.1130R} {493, 1130}

\bibitem[\protect\citeauthoryear{{Rauer} et~al.}{{Rauer}
  et~al.}{2014}]{2014ExA....38..249R}
{Rauer} H.,  et~al., 2014, \mn@doi [Experimental Astronomy]
  {10.1007/s10686-014-9383-4}, \href
  {https://ui.adsabs.harvard.edu/abs/2014ExA....38..249R} {38, 249}

\bibitem[\protect\citeauthoryear{{Rossiter}}{{Rossiter}}{1924}]{1924ApJ....60...15R}
{Rossiter} R.~A.,  1924, \mn@doi [\apj] {10.1086/142825}, \href
  {https://ui.adsabs.harvard.edu/abs/1924ApJ....60...15R} {60, 15}

\bibitem[\protect\citeauthoryear{{Ryabchikova}, {Piskunov}, {Kupka}  \&
  {Weiss}}{{Ryabchikova} et~al.}{1997}]{1997BaltA...6..244R}
{Ryabchikova} T.~A.,  {Piskunov} N.~E.,  {Kupka} F.,   {Weiss} W.~W.,  1997,
  \mn@doi [Baltic Astronomy] {10.1515/astro-1997-0216}, \href
  {https://ui.adsabs.harvard.edu/abs/1997BaltA...6..244R} {6, 244}

\bibitem[\protect\citeauthoryear{{Ryabchikova}, {Piskunov}, {Kurucz},
  {Stempels}, {Heiter}, {Pakhomov}  \& {Barklem}}{{Ryabchikova}
  et~al.}{2015}]{2015PhyS...90e4005R}
{Ryabchikova} T.,  {Piskunov} N.,  {Kurucz} R.~L.,  {Stempels} H.~C.,  {Heiter}
  U.,  {Pakhomov} Y.,   {Barklem} P.~S.,  2015, \mn@doi [\physscr]
  {10.1088/0031-8949/90/5/054005}, \href
  {https://ui.adsabs.harvard.edu/abs/2015PhyS...90e4005R} {90, 054005}

\bibitem[\protect\citeauthoryear{{Shahbaz} \& {Watson}}{{Shahbaz} \&
  {Watson}}{2007}]{2007A&A...474..969S}
{Shahbaz} T.,  {Watson} C.~A.,  2007, \mn@doi [\aap]
  {10.1051/0004-6361:20078251}, \href
  {https://ui.adsabs.harvard.edu/abs/2007A&A...474..969S} {474, 969}

\bibitem[\protect\citeauthoryear{{Simola}, {Dumusque}  \&
  {Cisewski-Kehe}}{{Simola} et~al.}{2019}]{2019A&A...622A.131S}
{Simola} U.,  {Dumusque} X.,   {Cisewski-Kehe} J.,  2019, \mn@doi [\aap]
  {10.1051/0004-6361/201833895}, \href
  {https://ui.adsabs.harvard.edu/abs/2019A&A...622A.131S} {622, A131}

\bibitem[\protect\citeauthoryear{{Simpson}, {Fetherolf}, {Kane}, {Li}, {Pepper}
   \& {Mo{\v{c}}nik}}{{Simpson} et~al.}{2022}]{2022AJ....163..215S}
{Simpson} E.~R.,  {Fetherolf} T.,  {Kane} S.~R.,  {Li} Z.,  {Pepper} J.,
  {Mo{\v{c}}nik} T.,  2022, \mn@doi [\aj] {10.3847/1538-3881/ac5d41}, \href
  {https://ui.adsabs.harvard.edu/abs/2022AJ....163..215S} {163, 215}

\bibitem[\protect\citeauthoryear{{Strassmeier} \& {Bopp}}{{Strassmeier} \&
  {Bopp}}{1992}]{1992A&A...259..183S}
{Strassmeier} K.~G.,  {Bopp} B.~W.,  1992, \aap, \href
  {https://ui.adsabs.harvard.edu/abs/1992A&A...259..183S} {259, 183}

\bibitem[\protect\citeauthoryear{{Thompson}, {Watson}, {de Mooij}  \&
  {Jess}}{{Thompson} et~al.}{2017}]{2017MNRAS.468L..16T}
{Thompson} A.~P.~G.,  {Watson} C.~A.,  {de Mooij} E.~J.~W.,   {Jess} D.~B.,
  2017, \mn@doi [\mnras] {10.1093/mnrasl/slx018}, \href
  {https://ui.adsabs.harvard.edu/abs/2017MNRAS.468L..16T} {468, L16}

\bibitem[\protect\citeauthoryear{{Triaud} et~al.,}{{Triaud}
  et~al.}{2009}]{2009A&A...506..377T}
{Triaud} A.~H.~M.~J.,  et~al., 2009, \mn@doi [\aap]
  {10.1051/0004-6361/200911897}, \href
  {https://ui.adsabs.harvard.edu/abs/2009A&A...506..377T} {506, 377}

\bibitem[\protect\citeauthoryear{{Valenti} \& {Piskunov}}{{Valenti} \&
  {Piskunov}}{1996}]{1996SME}
{Valenti} J.~A.,  {Piskunov} N.,  1996, \aaps, \href
  {https://ui.adsabs.harvard.edu/abs/1996A&AS..118..595V} {118, 595}

\bibitem[\protect\citeauthoryear{{Vogt} \& {Penrod}}{{Vogt} \&
  {Penrod}}{1983}]{1983PASP...95..565V}
{Vogt} S.~S.,  {Penrod} G.~D.,  1983, \mn@doi [\pasp] {10.1086/131208}, \href
  {https://ui.adsabs.harvard.edu/abs/1983PASP...95..565V} {95, 565}

\bibitem[\protect\citeauthoryear{{Vogt}, {Hatzes}, {Misch}  \&
  {Kurster}}{{Vogt} et~al.}{1997}]{1997astro.ph..4191V}
{Vogt} S.~S.,  {Hatzes} A.~P.,  {Misch} A.~A.,   {Kurster} M.,  1997, \mn@doi
  [arXiv e-prints] {10.48550/arXiv.astro-ph/9704191}, \href
  {https://ui.adsabs.harvard.edu/abs/1997astro.ph..4191V} {pp
  astro--ph/9704191}

\bibitem[\protect\citeauthoryear{{Wehrhahn}, {Piskunov}  \&
  {Ryabchikova}}{{Wehrhahn} et~al.}{2023}]{2023A&A...671A.171W}
{Wehrhahn} A.,  {Piskunov} N.,   {Ryabchikova} T.,  2023, \mn@doi [\aap]
  {10.1051/0004-6361/202244482}, \href
  {https://ui.adsabs.harvard.edu/abs/2023A&A...671A.171W} {671, A171}

\bibitem[\protect\citeauthoryear{{Wyttenbach}, {Ehrenreich}, {Lovis}, {Udry}
  \& {Pepe}}{{Wyttenbach} et~al.}{2015}]{2015A&A...577A..62W}
{Wyttenbach} A.,  {Ehrenreich} D.,  {Lovis} C.,  {Udry} S.,   {Pepe} F.,  2015,
  \mn@doi [\aap] {10.1051/0004-6361/201525729}, \href
  {https://ui.adsabs.harvard.edu/abs/2015A&A...577A..62W} {577, A62}

\bibitem[\protect\citeauthoryear{{de Leon} et~al.,}{{de Leon}
  et~al.}{2023}]{2023MNRAS.522..750D}
{de Leon} J.~P.,  et~al., 2023, \mn@doi [\mnras] {10.1093/mnras/stad894}, \href
  {https://ui.adsabs.harvard.edu/abs/2023MNRAS.522..750D} {522, 750}

\makeatother
\end{thebibliography}



\pagebreak
\appendix
\label{sec:appA}

\section{Feature in July 2007}
\label{sec:feature_sec}

\begin{figure*}
	\includegraphics[scale = 0.27, trim={0cm, 0, 7.9cm, 0}, clip]{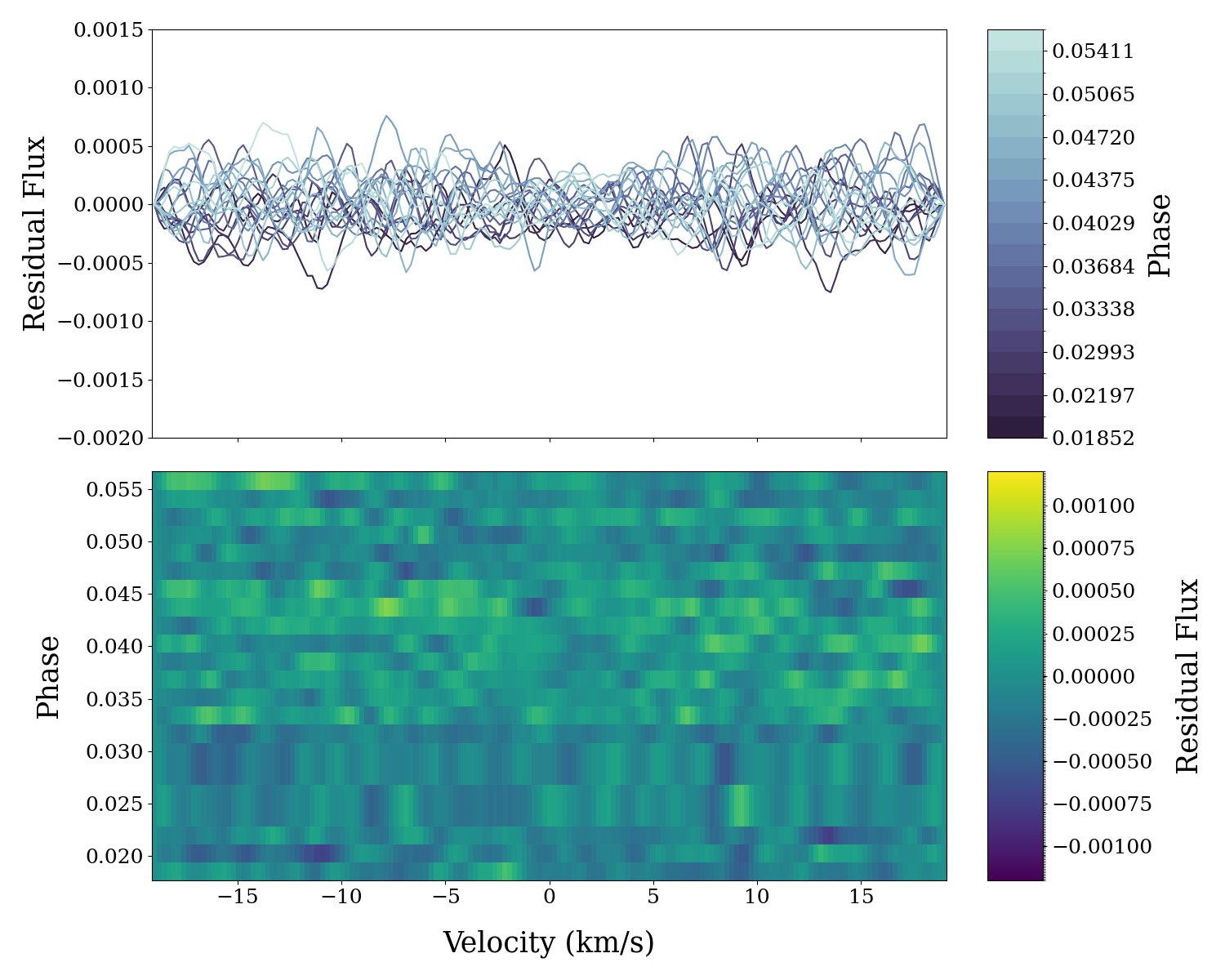} 
        \includegraphics[scale = 0.27, trim={4.5cm, 0, 0cm, 0}, clip]{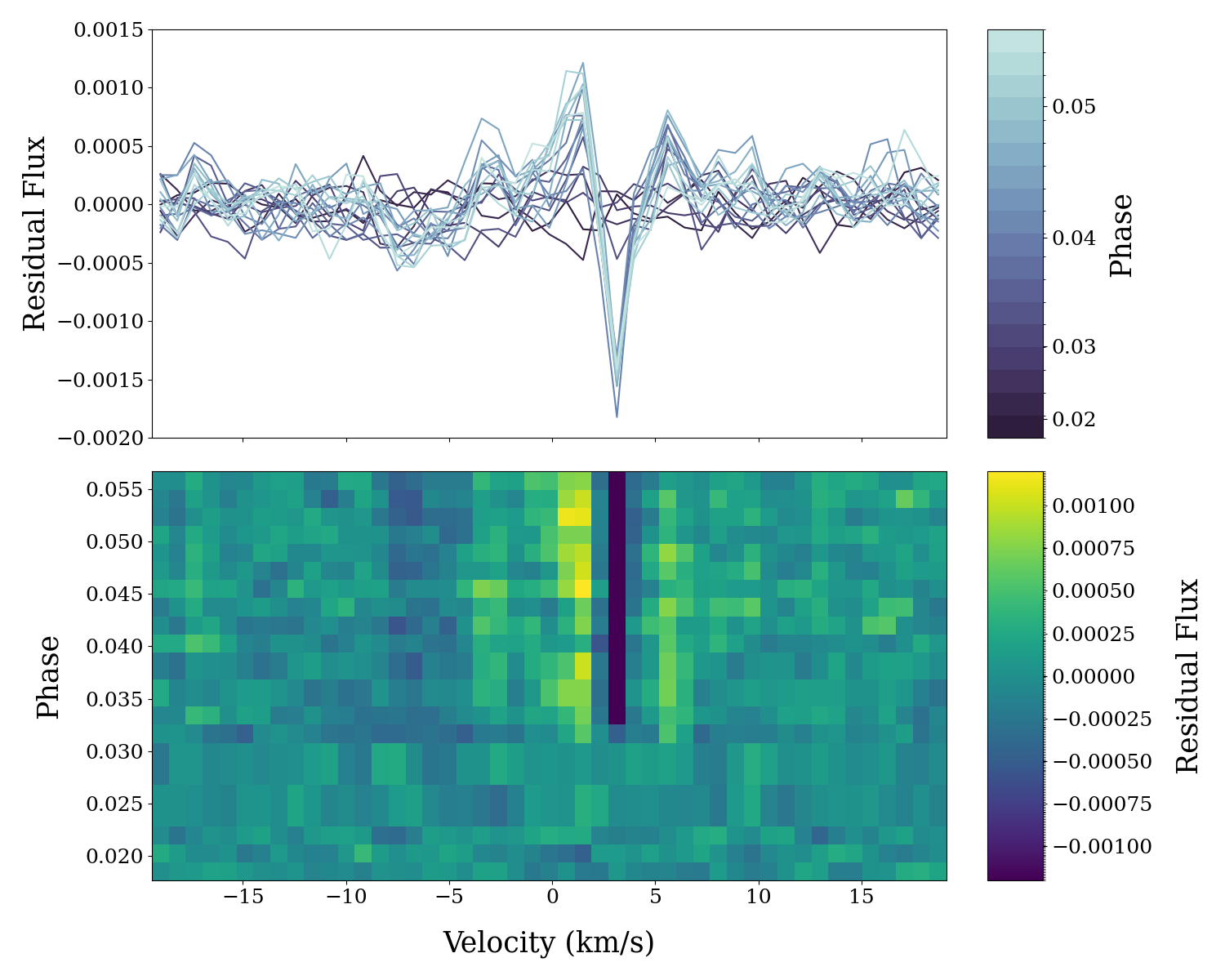}
    \caption{The out-of-transit residual profiles for July 2007, obtained via the Reloaded RM (RRM) method from HARPS CCFs (left) and $\codename$ profiles (right) are shown in the top panel for each case. Their corresponding phase maps are displayed in the bottom panel. A sharp decrease in flux (observed as a dark blue feature), surrounded by a flux increase (observed as a yellow feature) on either side can be seen at $\sim~3$\,km\,s$^{-1}$ in the residual $\codename$ profiles. This feature appears at a phase of $\sim0.032$ and is not present in the CCFs.}
    \label{fig:activity}
\end{figure*}

\begin{figure*}
    \includegraphics[scale = 0.4]{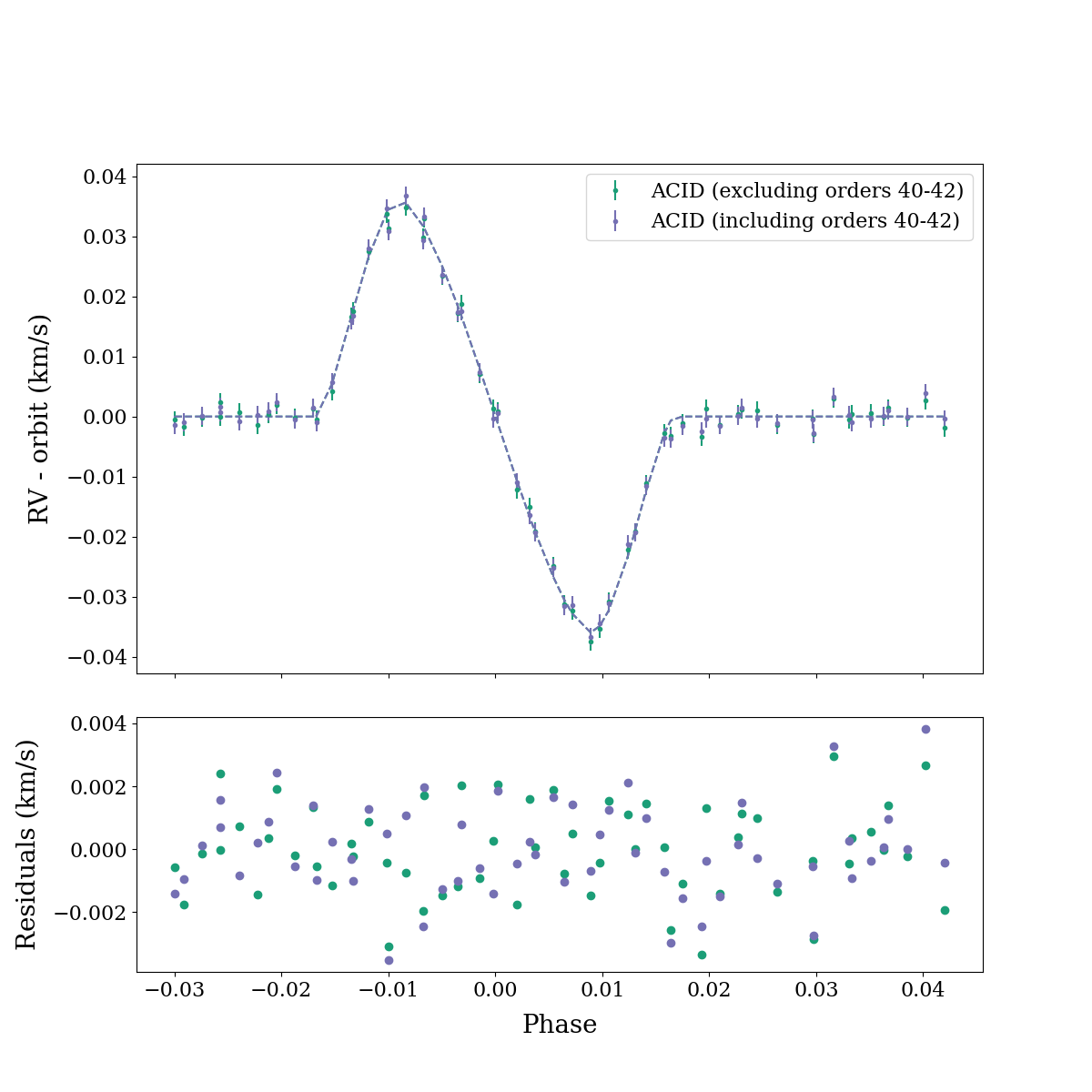}
    \centering
    \caption{Measured RVs from ACID (including orders 40-42) and ACID (excluding orders 40-42) with the best fit RM model indicated by a dashed line in each case.}
    \label{fig:app_rvs}
\end{figure*}

\indent It was seen that a feature was present in the residual $\codename$ profiles for July 2007. Fig.~\ref{fig:activity} shows the out-of-transit residual profiles with the residual CCFs on the left and the residual $\codename$ profiles on the right. In this case, the out-of-transit profile was taken as a weighted average of profiles with phase$\le$0.032. The feature can be seen as a sharp dip with a slight rise in residual flux on either side that stretches over several frames. This can be interpreted either as a feature present in the line profile from phase$\le$0.032 or that the profile becomes narrower beyond this phase. From further investigation, this feature was traced back to orders 40-42 (covering a wavelength range of $5028.603-5170.648$\AA) with the strongest signal being present in order 41. The feature is isolated to these orders only and no other orders show evidence of the feature. To further investigate the nature of this feature, $\codename$ was applied to the s1d spectra. This returned s1d residual profiles with no feature present hence indicating it is not of planetary or stellar origin. This may be therefore caused by instrumental effects that have been removed from the s1d spectra but not the e2ds. 
\newline \indent The RM analysis discussed in Section \ref{sec: RM_analysis} was repeated for ACID profiles including and excluding orders 40-42 for observations carried out in September 2006 and August 2007. Fig.~\ref{fig:app_rvs} shows the RVs measured, along with the best fit RM curve in each case. It can be seen that the RV measurements obtained agree within the uncertainties when including and excluding orders 40-42. Furthermore, the best fit RM model returns a projected spin-orbit angle of $\lambda = 0.13 \pm 0.34 \degree$ when including these orders, and $\lambda = 0.12 \pm 0.34 \degree$ when excluding them. As these results agree within the uncertainties we conclude that the feature is therefore not present in the other observation nights and is an anomalous result unique to July 2007. In addition, as this does not appear in other nights, or the s1d spectra, it cannot be an artifact produced from $\codename$ -- thus justifying the removal of these orders from July 2007 only.

\bsp	
\label{lastpage}
\end{document}